\documentclass[proof]{WileyASNA-v1}

\articletype{Article Type}%

\received{26 April 2016}
\revised{6 June 2016}
\accepted{6 June 2016}

\graphicspath{{./}{images/}}
\begin{document}

\title{Radial Metallicity Gradients for the Chemically Selected Galactic Thin Disc Main-Sequence Stars}

\author[1]{F. Akbaba*}
\author[2]{T. Ak}
\author[2]{S. Bilir}
\author[2]{O. Plevne}
\author[2]{\"{O}. \"{O}nal Ta\c s}
\author[3]{G. M. Seabroke}
\authormark{Akbaba \textsc{et al.}}

\address[1]{\orgdiv{Institute of Graduate Studies in Science}, \orgname{Istanbul University}, \orgaddress{\state{Istanbul}, \country{Türkiye}}}

\address[2]{\orgdiv{Faculty of Science, Department of Astronomy and Space Sciences}, \orgname{Istanbul University}, \orgaddress{\state{Istanbul}, \country{Türkiye}}}

\address[3]{\orgdiv{Mullard Space Science Laboratory, Holmbury St Mary, Dorking, Surrey, RH5 6NT}, \orgname{University College London}, \orgaddress{\state{London}, \country{United Kingdom}}}

\corres{*Furkan Akbaba, Faculty of Science, Department of Astronomy and Space Sciences, Istanbul, Türkiye. \email{furkan.akbaba@ogr.iu.edu.tr}}


\abstract{We present the radial metallicity gradients within the Galactic thin disc population through main-sequence stars selected on the chemical plane using GALAH DR3 accompanied with {\it Gaia} DR3 astrometric data. The [Fe/H], [$\alpha$/Fe] and [Mg/H] radial gradients are estimated for guiding radius as $-0.074\pm 0.006$, $+0.004\pm0.002$, $-0.074\pm0.006$ dex kpc$^{-1}$ and for the traceback early orbital radius as $-0.040\pm0.002$, $+0.003\pm 0.001$, $-0.039\pm 0.002$ dex kpc$^{-1}$ for 66,545 thin-disc stars, respectively. Alteration of the chemical structure within the Galactic disc caused by the radial orbital variations complicates results for the radial metallicity gradient. The effect of radial orbital variations on the metallicity gradients as a function on time indicates the following results: (i) The presence of a gradient along the disc throughout the time for which the model provides similar prediction, (ii) the radial orbital variations becomes more pronounced with the age of the stellar population and (iii) the effect of radial orbital variations on the metallicity gradients is minimal. The effect of radial orbital variations is found to be at most 6\% which does not statistically affect the radial gradient results. These findings contribute to a better understanding of the chemical evolution within the Galactic disc and provide an important basis for further research.
}


\fundingInfo{Türkiye Bilimsel ve Teknolojik Araştırma Kurumu, Grant/Award Number:119F072}

\maketitle


\section{Introduction}\label{sec1}

\label{introduction}
For centuries, astronomers have been fascinated by the Milky Way galaxy for its appearance in the night sky. With this inspiration, they are trying to understand its structure, formation and how it evolved to its current state by examining in depth the nearby stars. Understanding the mechanisms behind the formation of the Milky Way as a model for galaxy formation involves a long list of various models or their combinations. The study of Milky Way formation models has a long and rich history dating back to the early days of modern astronomy \citep{Matteucci12}. Our modern view of the Milky Way is founded on the early modelling studies of ‘Eggen-Lynden-Sandage” \citep[ELS,][]{ELS} and “Searle-Zinn” \citep[SZ,][]{SZ}. Over time, scientists' cumulative contributions to the perspectives of their predecessors have allowed formation models to gain a more realistic appearance with newly adapted statistical methods, developed with versatile observational data, and strengthened with computational capabilities. As a result of these efforts a new formation model that considers the chemistry of stars, the two-infall model of \citet{Chiappini1997} and its evolved version that considers the chemodynamical approach were designed \citet{Minchevetal2018}. These models have used pre-existing observational data so that they are tailored for the Milky Way, which gives them considerable predictive power. A full cosmological simulation of galaxy formation that involves chemical evolution with a galaxy with the same mass as the Milky Way also mimics the formation processes of the Galaxy \citep{Scannapieco2009}. To develop a new model or constrain existing Milky Way formation models, it is necessary to obtain the precise age, chemistry, kinematic and dynamic properties of the stars, and a detailed examination of the relationships and trends between these properties.

The metallicity gradients are valuable tools for revealing the structure, formation and evolution of the Milky Way. To determine the metallicity gradient in any direction, it is essential to determine precise distances. Various objects such as main-sequence stars \citep{Karaali2003, Ak2007a, Ak2007b, Yaz2010, Coskunoglu2011, Plevne2015, Guctekin2019}, red clump stars \citep{Bilir2012, OnalTas2016, OnalTas2018, Andersetal2017, Anders2023}, open clusters \citep{Magrinietal2009, CarreraPancino2011, Spinaetal2014, Carraroetal2016, Myersetal2022, Netopil2022}, and cepheids \citep{Genovalietal2014, Lemasleetal2013, Luck2018} have been employed for this purpose, as their distances can be precisely estimated. Metallicity gradients are generally calculated in the Galactocentric radial direction or vertically from the Galactic plane. Often, a metallicity gradient indicates the inside-out formation of the relevant part of the Galaxy, such as the thin disc \citep{MatteucciFrancois1989, Chiappini1997, Cescuttietal2007, Minchevetal2018, Sharmaetal2021}. However, \citet{SchonrichBinney2009} have demonstrated that an inside-out formation is not always necessary to produce a metallicity gradient. Regardless of the question of inside-out formation, metallicity gradient studies find their place in the literature as the most applied studies to test and develop the formation and evolution scenarios of the Galaxy.

Another important point of view in understanding the chemical evolution of the Milky Way lies in the choice of distance indicators for metallicity gradient determinations. As was emphasized in \citet{OnalTas2016}, there are two general approaches in these calculations. One depends on the static positions of stars inferred from their current Galactocentric positions, and the other depends on the dynamic position of stars derived from orbit calculations. Especially for the radial metallicity gradients, dynamically determined distances such as the mean Galactocentric distance, ($R_{\rm m}$) and the guiding radius of the stellar orbit, ($R_{\rm Guiding}$) come forward. Mean Galactocentric distance is derived from peri- and apo-Galactic distances of a stellar orbit, while guiding radius gives even more dynamic picture as the stellar orbit radially oscillates around it due to the effects of the large-scale perturbation sources within the Galactic potential field. In recent studies \citep[e.g.,][]{Ting2015, Khoperskov2020, Lian2022} that the guiding radius \citep{SchonrichBinney2009} is considered to be a key parameter related to radial migration and stellar chemistry. To do so, it helps recover lost information about the chemical evolution of the interstellar medium with cosmic time. On the other hand, the birth radius ($R_{\rm Birth}$) is more challenging to determine since it requires knowledge of the stellar age. 


The guiding radius is considered to be a key parameter as it relates to radial migration and stellar chemistry. Such that it helps to recover the lost information on the chemical evolution of the interstellar medium with cosmic time. Radial migration, specifically churning, significantly affects the interpretation of metallicity gradients, as stars move from their birth radii to their current positions over time, thus altering the gradients measured today from those at the time of star formation \citep[e.g.,][]{Pilkington2012, Minchev2013, Kubryk2013, Ratcliffe2023, Wang2024}. Therefore, it is crucial to consider the effects of radial migration when interpreting metallicity gradients to ensure accurate representations of the Galaxy's chemical evolution.

Moreover, understanding the birth radius is essential for gaining insights into the initial conditions of stellar populations and their subsequent evolution. Birth radius provides information about the original location where stars formed within the Galactic disc. This parameter is particularly useful when combined with age data, as it allows researchers to trace the movement of stars from their formation sites to their current positions. Simulation studies \citep[e.g.,][]{Andersetal2017, Carillo2023} have shown that stars can migrate significantly from their birth radii due to various dynamical processes, including interactions with spiral arms and giant molecular clouds. Therefore, incorporating $R_{\rm Birth}$ into analyses provides a more comprehensive understanding of the Milky Way’s chemical and dynamical evolution.

The Milky Way has been a challenging subject to study due to its complexity and ever-changing nature. Thanks to the surveys like {\it Gaia} \citep{GaiaDR3} and GALAH \citep{GalahDR3} have provided us with accurate data on more than half million stars in the Milky Way, which includes their positions, velocities, chemical compositions, and ages. Main-sequence stars have become an increasingly important tracer of the Milky Way's chemical evolution as they can be accurately dated using their luminosities and temperatures, enabling the reconstruction of their formation history and the chemical evolution of their parent clouds. The combination of {\it Gaia}'s precise astrometric measurements with GALAH's high-resolution spectroscopic data has opened up new opportunities for studying the Milky Way's main-sequence stars with unprecedented detail. The most recent catalogue, GALAH DR3 \citep{GalahDR3}, contains a total of 678,423 stars, of which 588,571 have been analyzed spectroscopically using the abundances of nearly 30 elements \citep{Buder2018}. The spectroscopic synthesis code Spectroscopy Made Easy (SME) \citep{PiskunovValenti2017} and CANNON \citep{Canon} are used to determine the atmospheric model parameters ($T_{\rm eff}$, $\log g$, [Fe/H], and $v_{\rm mic}$) and radial velocities of stars. 

The stellar age parameter is difficult to obtain because it is not a directly measured property. Its accuracy varies with the method and/or with the luminosity class of the objects. The isochrone fitting method is based on the Bayesian statistics of the observational priors defined from the stellar spectra. This application gives a probability density function \citep{PontEyer2004, JorgensenLindegren2005} which helps to define the most likely age. This method provides relatively precise ages \citep[$\pm$1 Gyr;][]{Edvardsson93, Casagrande11, Bensby14} for stars in a narrow spectral type (FGK) range, because it depends on the precision of stellar evolution tracks and only iron abundance ratios, so far. 

In this study, we not only present radial gradients for [Fe/H], $[\alpha/{\rm Fe]}$, and [Mg/H] metallicities but also investigate the implications of radial orbital variations for the chemodynamic structure of the Milky Way. Radial orbital variations, a key process influencing the metallicity distribution and the dynamics of stars within the Galaxy, challenges traditional models by altering the original chemical signatures of stellar populations. This phenomenon complicates the direct interpretation of metallicity gradients and necessitates a sophisticated analysis to disentangle the initial positions from the current locations of stars. In Section 2, we describe the properties of the spectroscopic data, applied selection criteria, and methods for determining the main-sequence star sample, astrometric data limitations, estimated distances, metallicity and stellar age calculation. In Section 3, the kinematic and Galactic orbital parameters of the programme stars and traceback early orbital radius calculations, and classification of Galactic populations are presented. Section 4 is dedicated to the radial metallicity gradients, model comparisons and incorporating the effects of radial orbital variations on the results. Finally, Section 5 includes a summary and conclusion.

\begin{figure}
\centering
\includegraphics[width=\columnwidth]{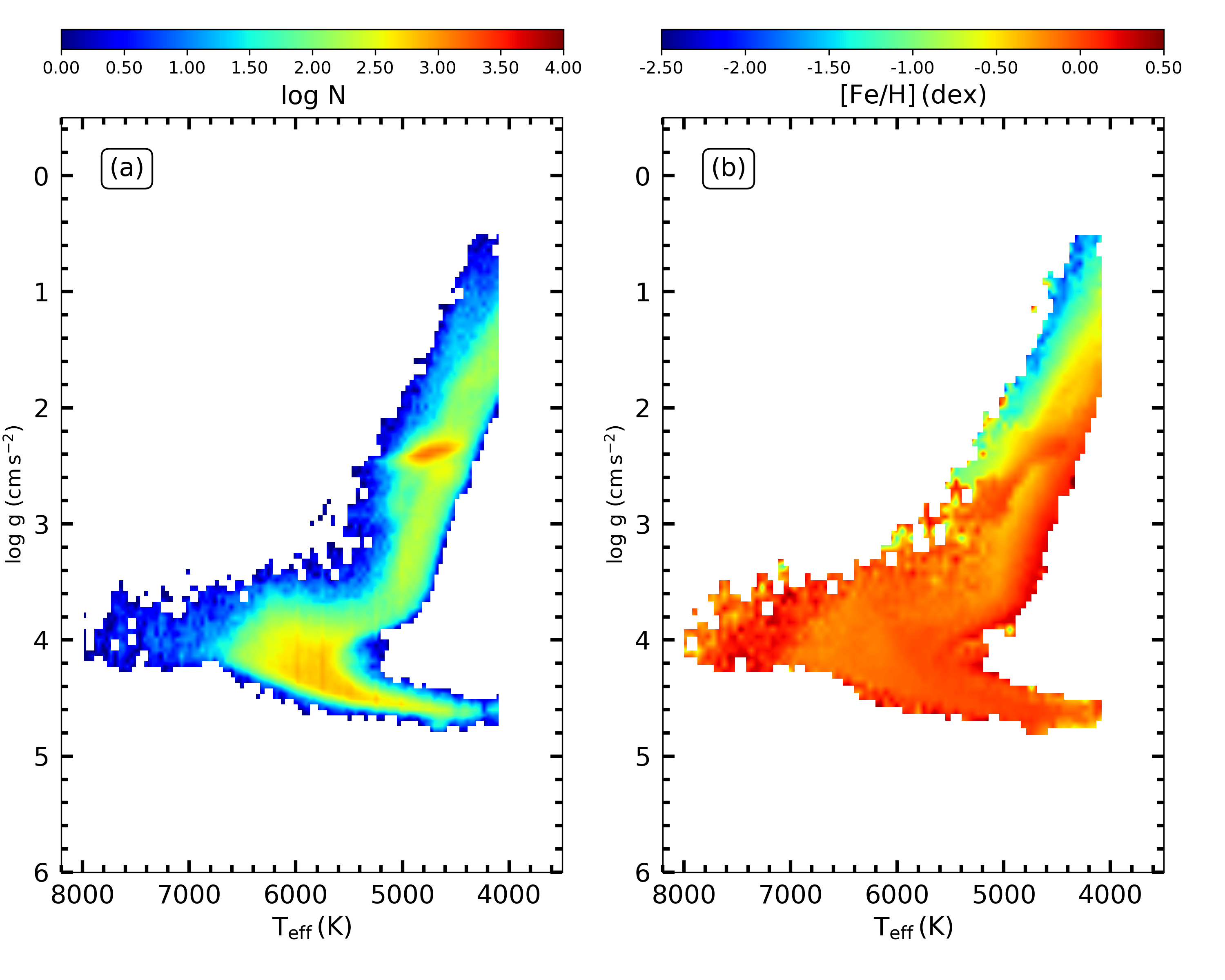}
\caption{Kiel diagrams of the 235,202 GALAH DR3 stars. Diagrams are colour-coded for the stellar number density (a) for the metallicity (b).}\label{fig:hrdiagrams}
\end{figure}

\begin{figure*}[ht!]
\centering
\includegraphics[width=\textwidth]{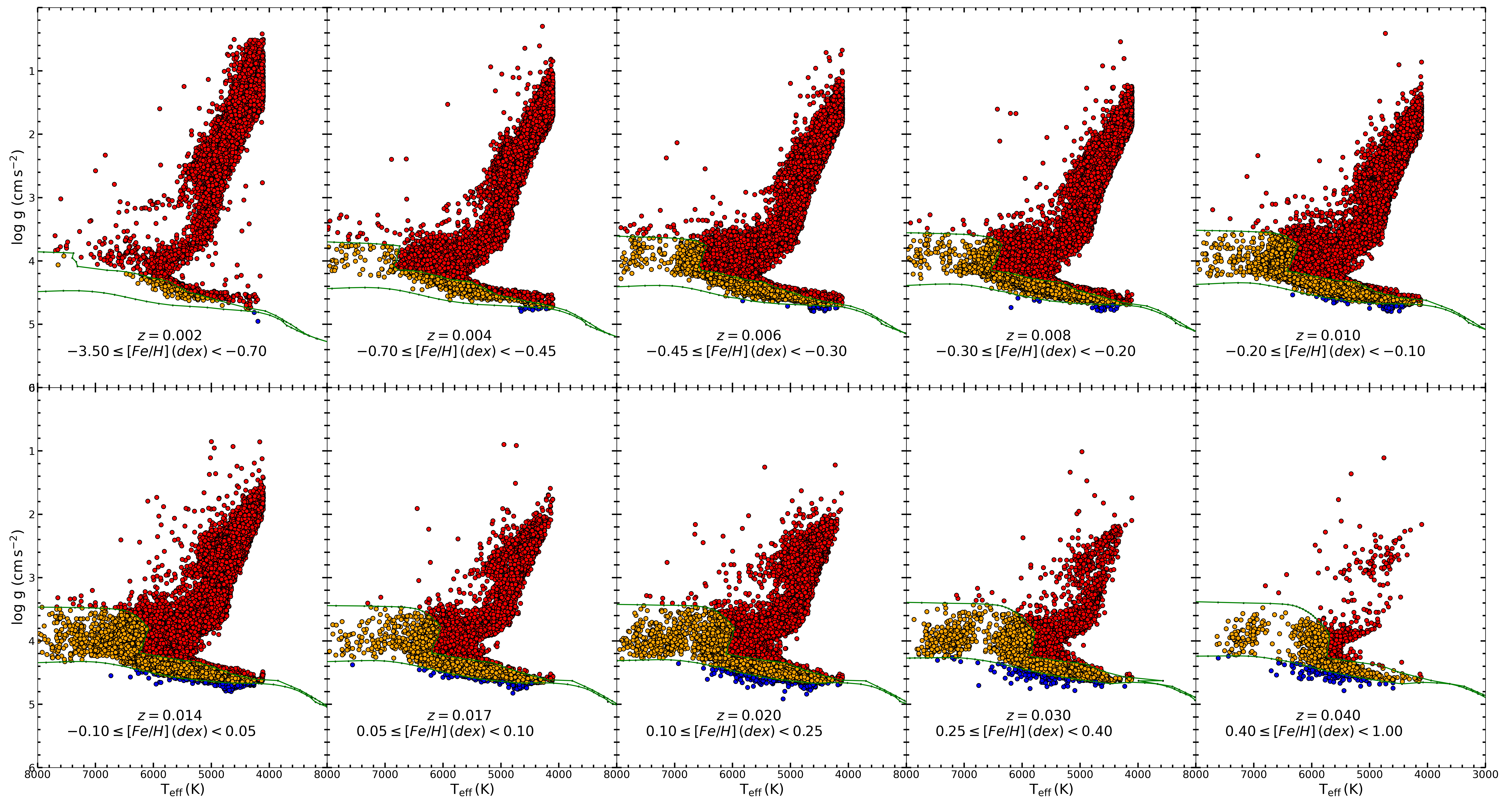}
\caption{Kiel diagrams of GALAH DR3 stars for 10 consecutive metallicity intervals in $-3.5\leq{\rm[Fe/H](dex})\leq1$. PARSEC ZAMS and TAMS curves are shown with green solid lines. Red, orange and blue dots represent evolved stars, main sequence stars and stars excluded from the statistical calculations, respectively. The mean heavy abundance ($z$) and metallicity ranges are given in the lower part of each Kiel diagram.}
\label{fig:ms_stars}
\end{figure*}

\section{Observational Data}
\label{observational data}
This study uses spectroscopic data from GALAH DR3 \citep{GalahDR3} and astrometric data from {\it Gaia} DR3 \citep{GaiaDR3}. There are 588,571 stars with equatorial ($\alpha$, $\delta$) and Galactic ($l$, $b$) coordinates, atmosphere model parameters ($T_{\rm eff}$, $\log~g$, [Fe/H], $\nu_{\rm mic}$), abundances of 27 elements, radial velocities ($V_{\rm R}$), proper-motion components ($\mu_{\alpha}\cos \delta, \mu_{\delta}$), trigonometric parallax ($\varpi$) and stellar age ($\tau$) from both catalogues. This study aims to investigate the chemodynamic structure of the Milky Way disc via metallicity gradients calculated for dynamic states of the Galaxy using selected main-sequence stars. In order to select main-sequence stars from the GALAH DR3 catalogue, several constraints are applied to the general catalogue. First, to minimize the spread of the data on the Kiel diagram, all stars with determined atmospheric model parameters, iron and alpha-element abundance ratios, radial velocity, and trigonometric parallax ($\varpi>0$ mas) were selected. Then, spectroscopic quality criteria suggested by the GALAH consortium are applied using spectral quality flags such as $flag\_sp=0$, $flag\_alpha\_fe=0$, $flag\_fe\_h=0$ and [Fe/H]$_{\rm err}\leq0.10$ dex. Additionally, stars with calculated ages are included in these data constraints. After these, the sample is reduced to 281,143 stars. The Kiel diagrams of the sample are shown in Figure~\ref{fig:hrdiagrams} with panels (a) and (b), which correspond to colour coding based on the number density and metallicity, respectively. The mean of the mode values of $S/N$ for four CCD filters, $S/N>35$ is applied as a constraint to the sample quality cut also suggested by the GALAH consortium. Following all these quality limitations, the final sample is reduced to 235,202 stars.

Main-sequence stars are selected with the method of \citet{Biliretal2020}. In this method, Kiel diagrams of all GALAH DR3 sample is drawn for 10 consecutive metallicity subsamples from -3.5 dex to 1 dex. In order to determine the luminosity classes, zero-age main sequence (ZAMS) and terminal age main-sequence (TAMS) curves, which are determined from PARSEC mass tracks \citep{Bressan2012}, are fitted to the Kiel diagram of each metallicity subsample. Kiel diagrams of each metallicity bin are shown in Figure~\ref{fig:ms_stars} with 10 panels. ZAMS and TAMS curves are represented with solid green lines. The ZAMS curve is below the TAMS one. Stars remaining between the ZAMS and the TAMS curves are accepted as main-sequence stars (yellow circles), while stars above the TAMS curve are accepted as evolved stars (red circles) like sub-giant ($\log g>3.5$ cgs), red giant branch, and red clump ($\log g<3.5$ cgs). Results of this analysis have shown that there are 75,698 main sequence and 158,770 evolved stars out of 235,202 stars. On the other hand, 734 stars below the ZAMS curve, which have no assigned luminosity class, are excluded from the statistics.

\begin{figure}
    \centering
    \includegraphics[width=\columnwidth]{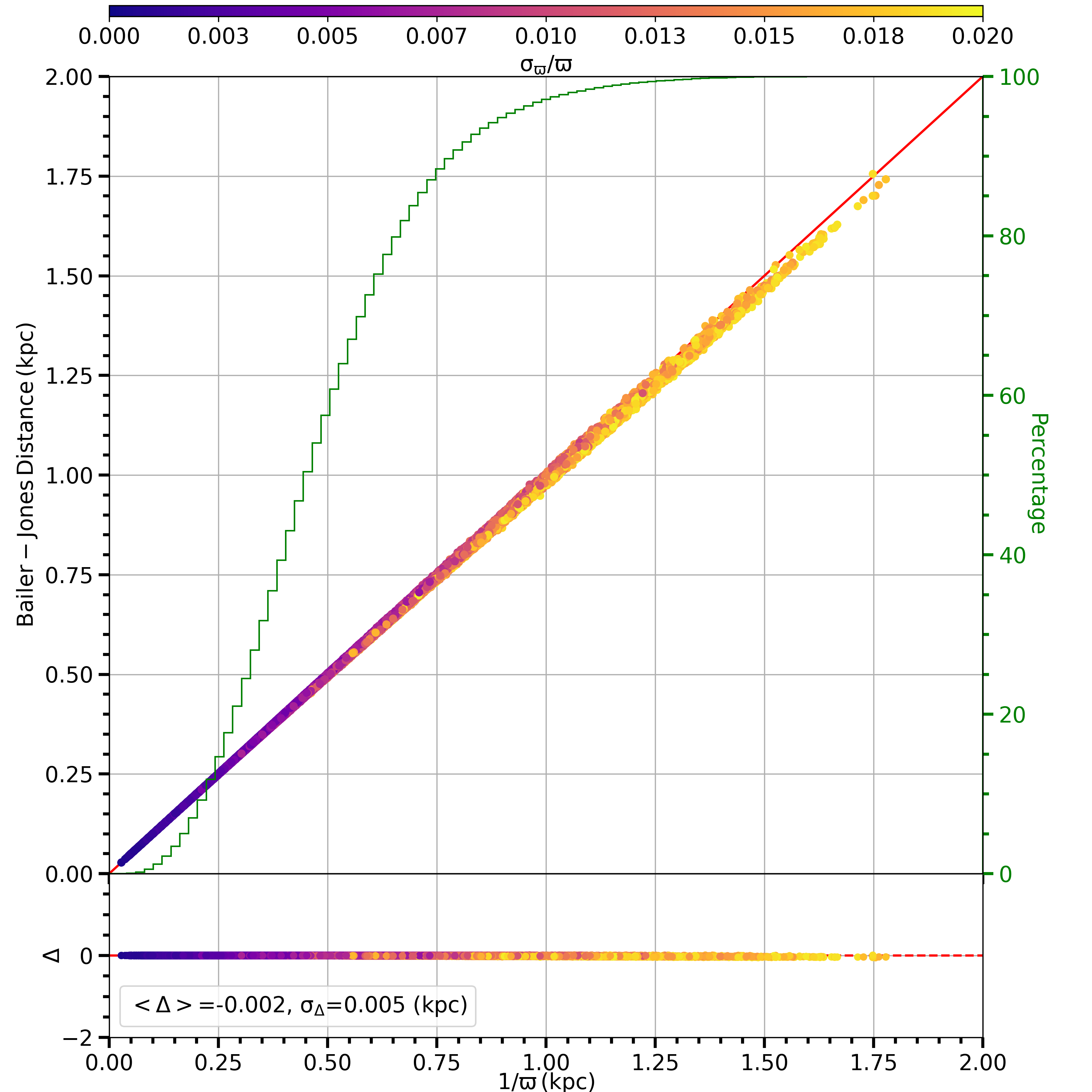}
    \caption{Comparison of the inverse parallax distance values for the sample with those obtained distance from the study by \citet{Bailer21}. The colouring is based on the relative parallax error.}
    \label{fig:dist-comp}
\end{figure}

Astrometric data of the main-sequence stars are provided from {\it Gaia} DR3 \citep{GaiaDR3} catalogue, which is pre-matched by the GALAH consortium \citep{GalahDR3}. Stellar distances are calculated via $d$(pc)=1000/$\varpi$(mas) relation using {\it Gaia} DR3 trigonometric parallaxes. The uncertainties on stellar distances are determined from the relative trigonometric parallax errors that are less than $\sigma_{\varpi}/\varpi \leq 0.05$ for the whole sample. We put $\sigma_{\varpi}/\varpi \leq 0.02$ limit to minimize the errors caused by the distances in kinematic and Galactic orbit parameter calculations. This limit represents 96.5\% of the main-sequence sample. After this limitation, our sample size was reduced to 73,048. This sample reaches 1.6 kpc with a median distance of 400 pc. These distances are compared with the Bayesian analysis-based distance calculation of \citet{Bailer21} by taking into account the zero point values of \citet{Lindegren}. In Figure \ref{fig:dist-comp}, we have shown the comparison of the distances and their residuals for sample main-sequence stars with colour code for relative trigonometric parallax errors. It is easily inferred from the Figure~\ref{fig:dist-comp}, that there is no significant change in difference calculations up to 1.5 kpc distance. However, the sample fills its potential at 1.25 kpc. So, we have chosen the conventional method of distance calculation.

\begin{figure}
\centering
\includegraphics[trim={0 3cm 0 3cm}, clip, width=\columnwidth]{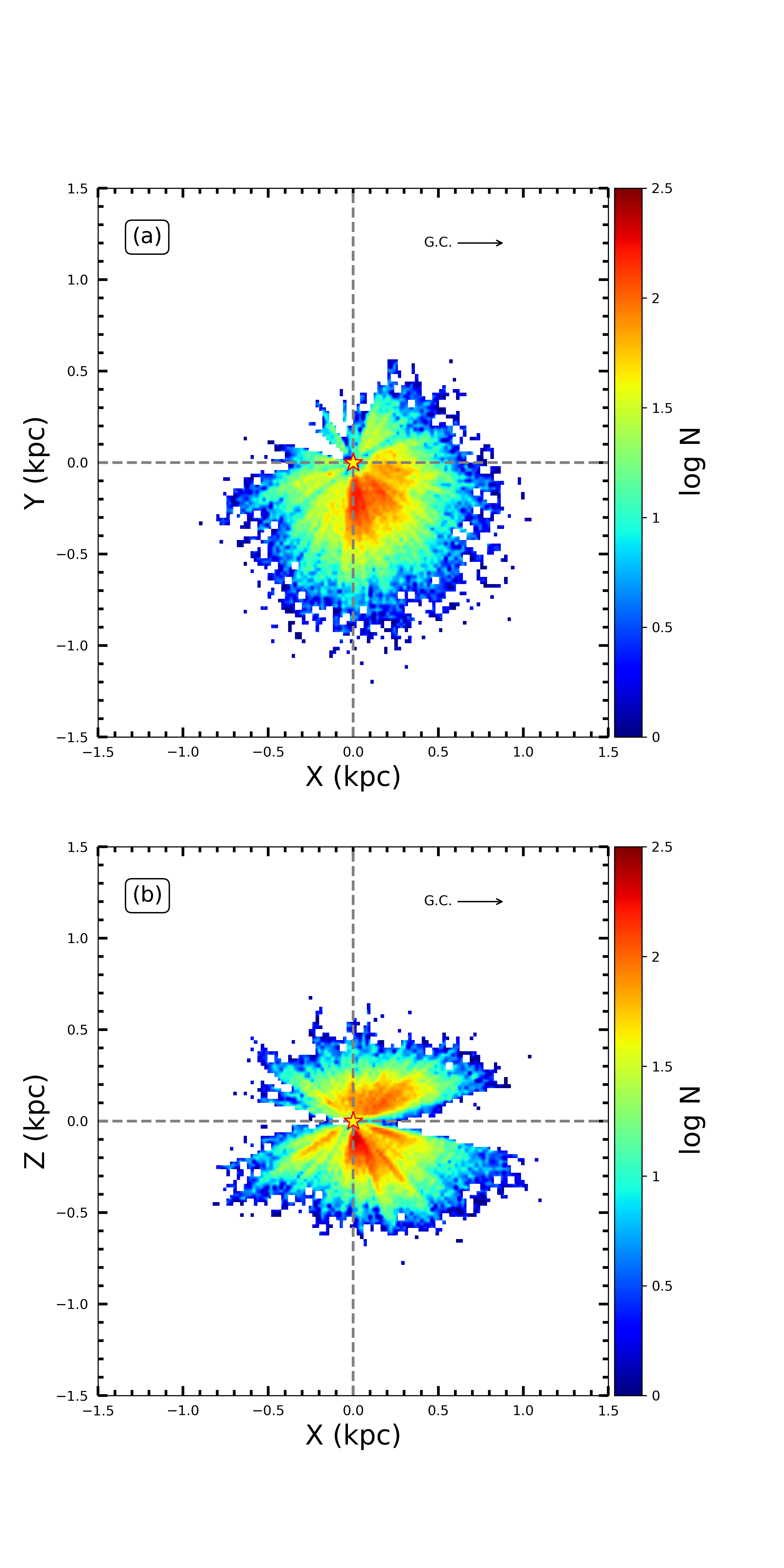}
\caption{Distribution of stars in heliocentric rectangular Galactic coordinate planes: on $Y \times X$, (a) and on $Z \times X$ (b). Both diagrams are colour-coded for stellar number density. Gray dashed vertical and horizontal lines denote 0 kpc and their intersection gives the position of the Sun which is represented with a star symbol.} \label{fig:xyxzdist}
\end{figure}

To study the spatial distributions of the main-sequence star sample, we calculated the heliocentric rectangular Galactic coordinates ($X$ towards the Galactic center, $Y$ Galactic rotation, $Z$ north Galactic Pole). The distributions of the sample in the $Y\times X$ and $Z \times X$ planes are colour-coded in the logarithmic number density of the stars in panels (a) and (b) of  Figure~\ref{fig:xyxzdist}, respectively. Most of the sample stars reside on the third and the fourth Galactic quadrant and the low number of stars are on the second quadrant. The stars reached 1 kpc on $X$ and $Y$, and 0.6 kpc on $Z$. Median distances on $X$, $Y$, and $Z$ are 100, -210, and -100 pc, respectively. 

Variation of the total proper motion values of the stars varies from -80 to +80 mas yr$^{-1}$. The calculated median value of the total errors of the proper motions is 0.02 mas yr$^{-1}$. The radial velocities vary in $-100< V_{\rm R}~({\rm km~s^{-1}})<110$ interval with the median value of 6.53 km~s$^{-1}$. This positive value reflects of the Galactic quadrants of the main-sequence sample, which is dominated by the stars from the third and the fourth Galactic quadrants. Uncertainty of the radial velocities ranges between 0.1 to 0.6 km s$^{-1}$ and their median is 0.12 km s$^{-1}$.

\subsection{Metallicity and Stellar Age}

Testing the structure of the Galactic disc with chemo-dynamic models requires well-known stellar distances and metallicities of the Galactic objects. For this study, both parameters are provided precisely from the state-of-the-art spectroscopic and astrometric surveys, i.e., GALAH DR3 and {\it Gaia} DR3. Iron and  $\alpha$-element\footnote{$[\alpha/{\rm Fe}] = ([{\rm Mg/Fe}] + [{\rm Si/Fe}] + [{\rm Ca/Fe}] + [{\rm Ti/Fe}])/4$} abundances of sample stars vary in $-1<{\rm [Fe/H] (dex)}<0.6$ and  $-0.4<[\alpha/{\rm Fe] (dex)}<0.4$, respectively. The median and standard deviation values of these parameters are $-0.04\pm0.22$ dex and $0.02\pm0.08$ dex, respectively. Apart from the $\alpha$-element abundance ratio, magnesium (Mg) element is considered alone in this study. Mg is produced via the carbon-burning stage in stellar evolution and it is returned to the interstellar medium with supernova type II. It also plays a crucial role in tracing the chemical evolution of the Galactic disc, as [Mg/Fe] or [Mg/H] ratios are fundamental in distinguishing between the high- and low-$\alpha$ disc populations \citep{Palla2022}. The [Mg/H] abundances of the main-sequence stars in the sample were found to be in the range $-0.65<{\rm [Mg/H] (dex)}<0.75$.
 
\begin{figure}[!ht]
    \centering
    \includegraphics[width=\columnwidth]{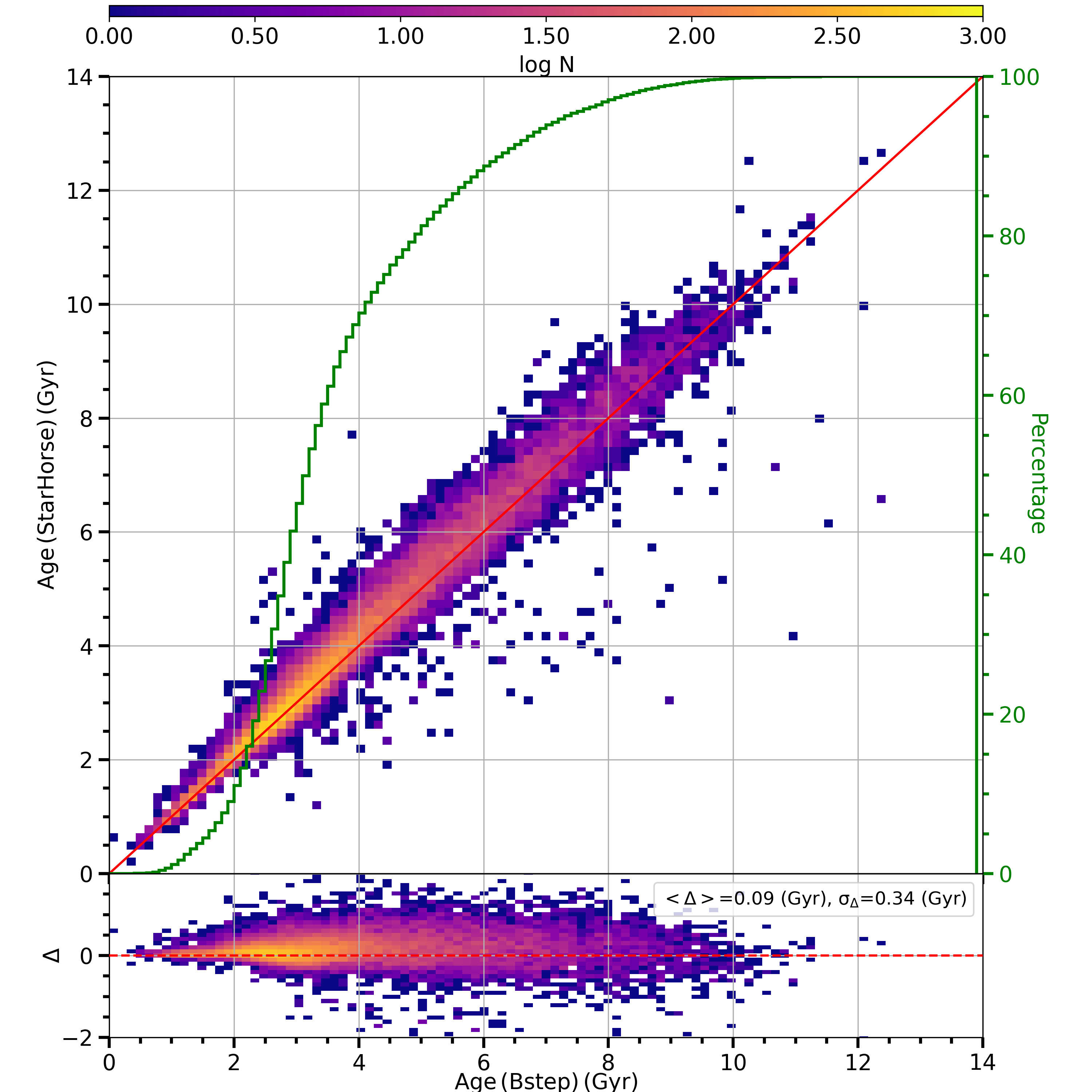}
    \caption{Comparison of the ages of sample stars calculated in the BSTEP \citep{Sharmaetal2018} with those calculated in the StarHorse catalogue \citep{StarHorse}. The colouring is based on the stellar number density.}
    \label{fig:age-comp}
\end{figure}

Stellar age is a pre-calculated parameter of the GALAH DR3 catalogue, which was calculated using the Bayesian Stellar Parameter Estimation (BSTEP) code, which implements the calculation technique of \citet{Sharmaetal2018}. The BSTEP code compares the observational parameters and PARSEC isochrones \citep{Bressan2012} in order to determine the properties like age, mass, and distance precisely. The stellar age of the sample stars ranges in $0<\tau {\rm (Gyr)}<13.5$, and their median age with its standard deviation is $5.39\pm 2.75$ Gyr. There exists another powerful tool known as StarHorse that applies the Bayesian statistics in the literature \citep{StarHorse}. A comparison between these age estimates and their residuals for sample stars is presented in Figure \ref{fig:age-comp}. The analyses showed that the mean difference between the two data sets was 0.09 Gyr, with a standard deviation of 0.34 Gyr. 

\section{Methods}
\subsection{Space Velocity and Orbital Calculations}

\begin{figure}
\centering
\includegraphics[width=\columnwidth]{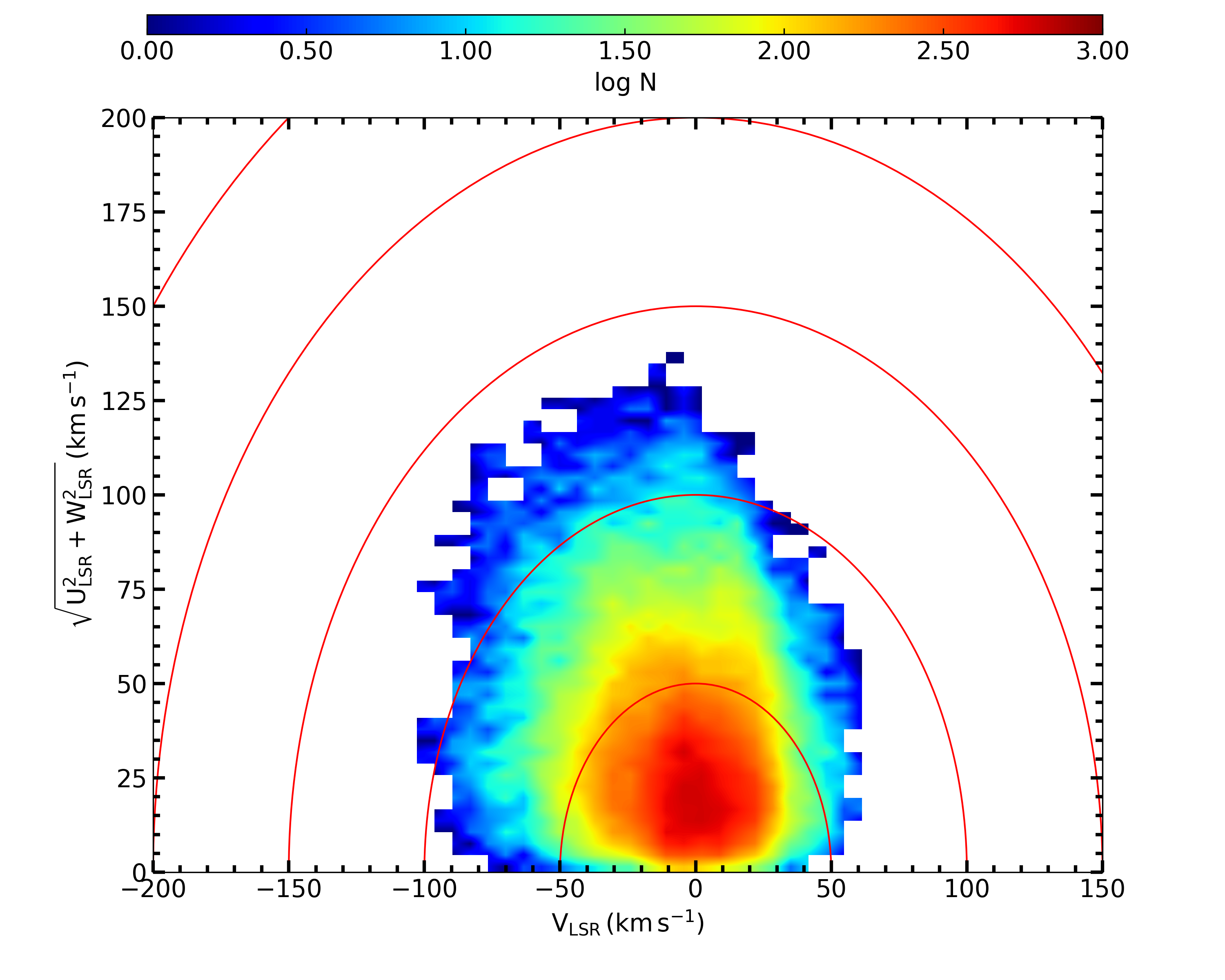}
\caption{Toomre diagram of sample stars. Red curves represent 50, 100, 150 and 200 km s$^{-1}$ iso-velocity circles.} \label{fig:toomren}
\end{figure}

Space velocity components, their respective uncertainties, and the Galactic orbital parameters are calculated using the {\it galpy} library \citep{galpy}. All these calculations are done by using the astrometric data (equatorial coordinates, proper-motion components, and trigonometric parallaxes) of {\it Gaia} DR3 and spectroscopic data (radial velocity) of GALAH DR3. Uncertainty analysis on the total space velocity component ($S_{\rm err}$) is important because, due to its spread, another constraint might apply to improve the quality of the kinematic properties of the main-sequence sample. Our analysis shows that total space velocity error varies in $0<S_{\rm err}\,{\rm (km\,s^{-1})}<1.8$ range with a median of 0.24 ${\rm km\,s^{-1}}$ and with a standard deviation of 0.23 ${\rm km\,s^{-1}}$. These descriptive statistics show the errors in space velocity components are small and precise. Thus, no extra constraint is applied to the main-sequence star sample. On the other hand, space velocity components contain two biases. One is due to differential rotation, and the other one is related to heliocentric analysis in the Solar neighbourhood. The differential rotation affects the space velocity components parallel to the Galactic disc by causing a lag on $U$ and $V$ space velocity components. To correct these velocities, $dU$ and $dV$ corrections are calculated for each velocity component of each main-sequence star in the sample using the method of \citet{MihalasBinney1981}. These correction equations consider both line-of-sight and tangential velocities along the Galactic longitude of the sample star. $dU$ and $dV$ velocity corrections are subtracted from the individual $U$ and $V$ velocities, respectively. Variation of each differential correction change in $-47.34 \leq dU\,{\rm (km\,s^{-1})} \leq 21.21$ and $-2.63 \leq dV\,{\rm (km\,s^{-1})} \leq 3.14$ ranges. The second bias is known as the Local Standard of Rest (LSR) correction. Spectroscopic observations are performed in the heliocentric system, and this cause in each ($U, V, W$) space velocity component contains the Sun's $(U, V, W)_{\odot}$. Thus, subtracting the already defined Solar space velocity components from each star's space velocity components cancels the bias. In this study, we have used \citet{Coskunoglu2011} values for the Sun's $(U, V, W)_{\odot}=(8.83\pm0.24,14.19\pm0.34, 6.57\pm0.21)$ ${\rm km\,s^{-1}}$. After these bias corrections, the distribution of the space velocities on the Toomre diagram is shown in Figure~\ref{fig:toomren} with colour coding based on the logarithmic number density of the main-sequence sample. 

\begin{figure*}[hbt!]
\centering
\includegraphics[width=\textwidth]{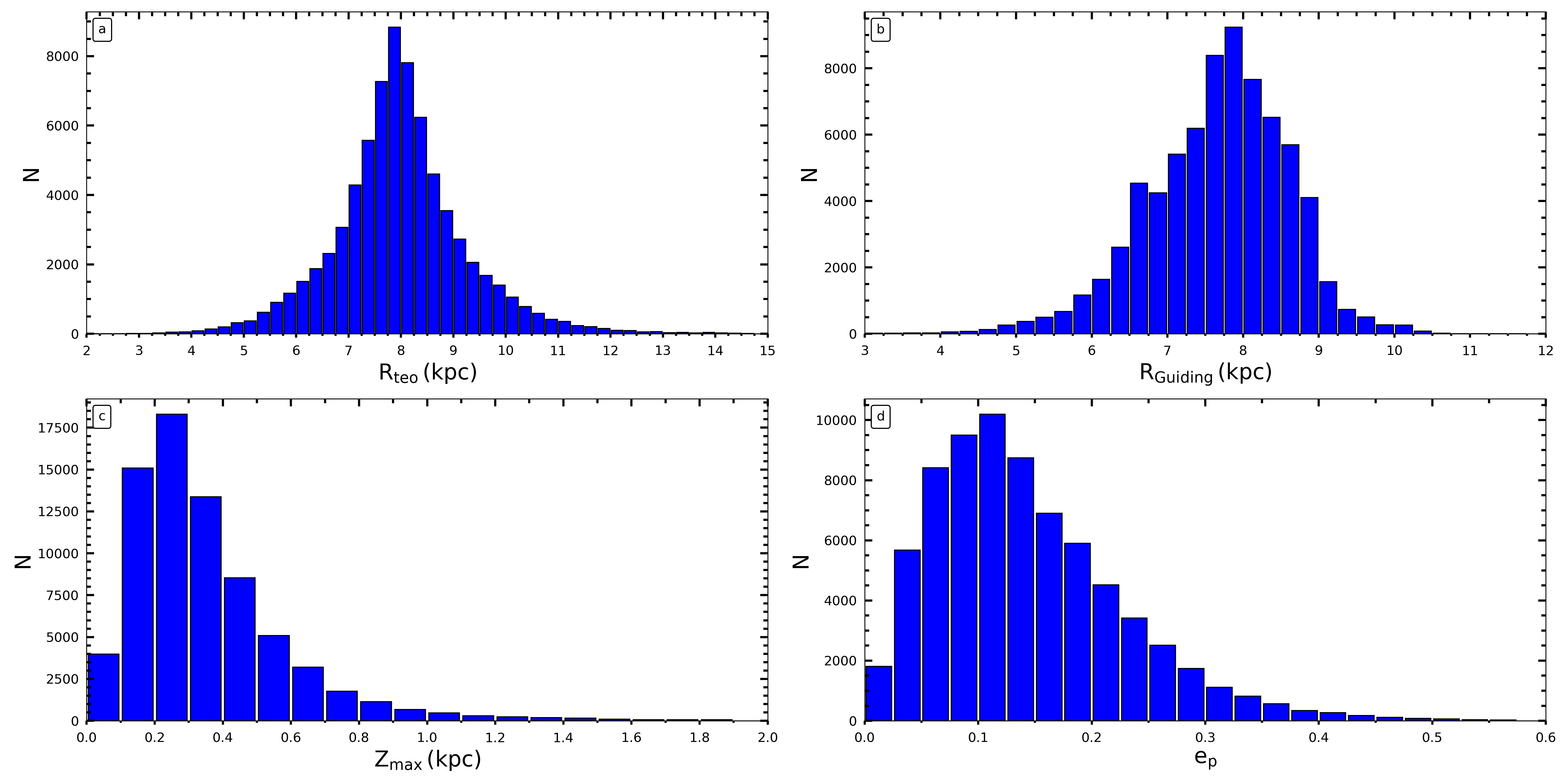}
\caption{Histograms of basic orbital parameters calculated for axisymmetrical Galactic potential. Panels are traceback early orbital radii (a), guiding radii (b), maximum vertical distance from the Galactic plane (c) and planar orbital eccentricity (d).} \label{fig:starsorbits}
\end{figure*}

Galactic orbit parameters are calculated using the {\sc MWPotential2014} model from \citet{galpy} {\it galpy} library. This model is an axisymmetrical model of the Milky Way that does not consider the effects of large-scale perturbation sources such as the Galactic bar, spiral arms, or giant molecular clouds. The model is a combination of analytic functions that describe the Milky Way. The Galactic bulge is modelled with Hernquist's potential; the disc is modelled with Miyamoto-Nagai potential, and the halo is modelled with The Navarro–Frenk–White potential \citep[for further details, see][]{galpy}. The model has produced the orbital parameters with steps of 10,000 years for 5 Gyr; with this long, a star can close its orbit. By doing so, it allows us to obtain the most precise orbital parameters possible. The output model parameters are peri-Galactic ($R_{\rm p}$) and apo-Galactic distances ($R_{\rm a}$), and the maximum vertical distance from the Galactic plane ($Z_{\rm max}$). Moreover, the guiding radius is calculated with the $R_{\rm Guiding} = r\times(V_{\phi}/V_{\rm c})$ relation, where $r$ is the radius of the Galactic orbit of the star, $V_{\phi}$ and $V_{\rm c}$ are tangential and circular velocity with respect to the Galactic center of the stellar orbit \citep{Minchevetal2018}. The orbital eccentricities of the stars are calculated using the following equation as $e_{\rm p}=(R_{\rm a}-R_{\rm p})/(R_{\rm a}+R_{\rm p})$. Histograms of the orbital parameters are shown in Figure~\ref{fig:starsorbits} with four panels. 
  
\subsubsection{Traceback Early Orbital Radius}

Understanding the early orbital characteristics of stars is essential for exploring their initial conditions and subsequent evolution within the Galactic disc. In this study, we developed a method to calculate what we term the traceback early orbital radius ($R_{\rm teo}$), which approximates a star’s early orbital configuration by tracing its orbit backwards in time under the assumption of a static Galactic potential. This calculation offers valuable insights into the early distribution of stars in the Galactic disc and how orbital variations have reshaped this distribution over time, contributing to our understanding of the Milky Way’s chemical and dynamical evolution \citep{Minchevetal2018}.

The calculation of $R_{\rm teo}$ involves integrating stellar orbits backwards in time using their current positions, velocities, and ages as initial conditions. This approach has been applied in previous studies \citep[e.g.,][]{Koc2022K, Yontan2022, Yontan2023b, Yontan2023, Tasdemir2023, Dursun2024, Yucel2024} to reconstruct the orbital characteristics of stars in the Milky Way.

In our method, we use the same orbital integration technique but extend its application by focusing on early orbital positions rather than attempting to estimate precise birth locations. Literature methods to determine the birth radius often rely on modelling the metallicity gradient as a function of time and correlating it with stellar ages and metallicities. These approaches assume a steady, linear metallicity gradient across Galactic radii and epochs \citep{Minchevetal2018, Prantzosetal2018, Prantzosetal2023, Ratcliffeetal2023}. While effective, they depend heavily on accurate age estimates and consistent metallicity trends across the Galactic disc, which can be influenced by radial migration and chemical homogeneity assumptions. In contrast, our method calculates $R_{\rm teo}$ by tracing stellar orbits backwards under a static Galactic potential, without requiring assumptions about metallicity gradients. The $R_{\rm teo}$ represents early orbital characteristics rather than a true birth radius. This approach minimizes reliance on external metallicity trends, thereby reducing sensitivity to systematic biases caused by variations in metallicity gradients across different Galactic regions and times. It provides a complementary, chemical evaluation model-independent perspective on the stars’ early orbital distributions within the disc.

In summary, our study emphasizes that $R_{\rm teo}$ is calculated independently of any assumed metallicity gradient but under the assumption that the Galactic potential remains constant over time. While this assumption simplifies the calculations, it introduces some uncertainty, as it neglects the effects of perturbative sources such as spiral arms and the Galactic bar. These perturbations can alter stellar orbits over time, leading to uncertainties that are not directly measurable due to the complexity of the dynamical processes. Nevertheless, as the primary aim of this study is to investigate the metallicity gradient of the disc and its variation over time, the $R_{\rm teo}$ approach was specifically designed to avoid reliance on metallicity gradient assumptions, providing a robust framework for this analysis.

\subsection{Classification of Galactic Populations} \label{sec:cgp}

The concept of determining Galactic chemical evolution through the chemical distribution in stars originated with the study by \citet{GilmoreWyseKuijen1989}, which demonstrated that stars in different Galactic populations exhibit distinct [O/${\rm Fe}]$ and [Fe/H] ratios. This observation led to the idea of chemical tagging \citep{FreemanBlanHawthor2002}, a powerful tool for identifying Galactic populations from any stellar sample with known iron and $\alpha$-element abundance ratios. In this study, we separated sample stars into Galactic populations on the $[\alpha/{\rm Fe}]\times{\rm [Fe/H]}$ plane using the population separation line determined by \citet{Plevne2020}. The separation line by \citet{Plevne2020} serves as a decision boundary on the chemical plane between two Gaussian Mixture Models (GMMs), one representing the low-$\alpha$ population (i.e., thin disc) and the other high-$\alpha$ (i.e., thick disc or halo) population. GMM is an unsupervised machine learning algorithm that classifies data by fitting the preferred number of Gaussian distributions and calculating the probabilities of each. The separation line represents the region of equal probability between the two populations.

\begin{figure}
\centering
\includegraphics[width=\columnwidth]{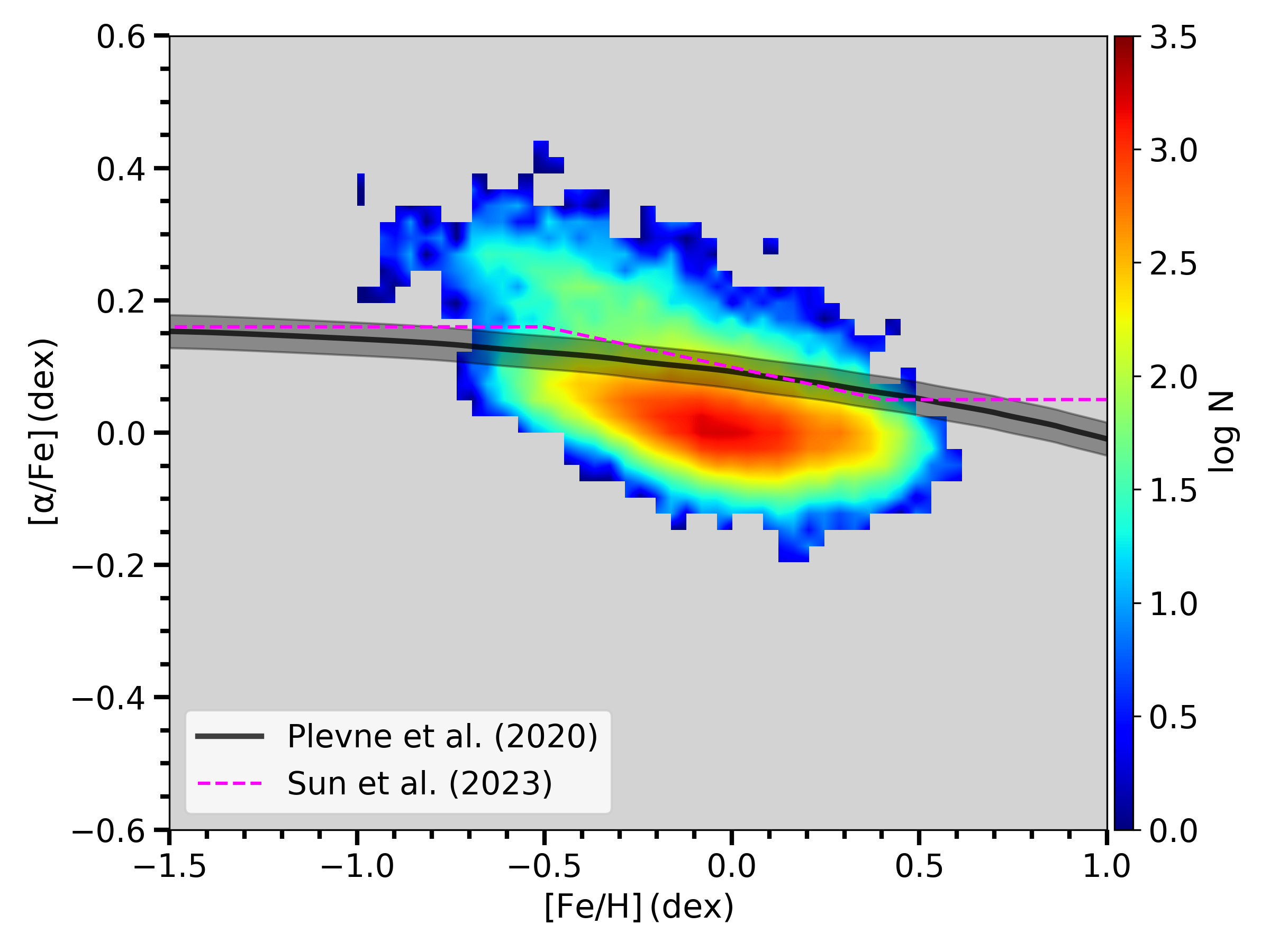}
\caption{Positions of sample stars on $[\alpha/{\rm Fe}]\times$[Fe/H] plane. Diagrams colour-coded for logarithmic number density. Gray dashed curve represents the decision boundary line of the GMM.} \label{fig:galpopn}
\end{figure}

Figure~\ref{fig:galpopn} shows that the main-sequence sample lacks a distinct high-$\alpha$ population, while the low-$\alpha$ population is clearly identifiable. In this study, we did not recalculate the boundary line as in \citet{Plevne2020}; instead, we used the line as a classifier and compared it with the ones from \citet{Sunetal2023}. In Figure~\ref{fig:galpopn}, the separation line from \citet{Sunetal2023} is represented by a magenta dashed line, while the line from \citet{Plevne2020} is shown as a solid black line. The black-shaded region indicates the 95\% confidence region for the boundary line. The figure demonstrates that the two lines are very similar. Using the \citet{Plevne2020} line as a classifier, our sample separation resulted in 66,545 stars belonging to the low-$\alpha$ population and 6,503 stars to the high-$\alpha$ population. When using the \citet{Sunetal2023} line, the results were 67,727 stars in the low-$\alpha$ population and 5,321 in the high-$\alpha$ population. Although the results were similar, we chose to use the \citet{Plevne2020} line because it suggested a greater number of high-$\alpha$ populations in the sample. This choice allowed us to conduct the study with a less contaminated low-$\alpha$ population. It is also found that about 97\% of the low-$\alpha$ population separated in the $[\alpha/{\rm Fe}]\ \times$[Fe/H] plane is located at $Z\leq825$ pc distances \citep{Ak2015}. This finding supports that the low-$\alpha$ population separation based on both chemical and dynamical orbital parameters is very accurate.

\begin{figure*}[ht!]
\centering
\includegraphics[width=\textwidth]{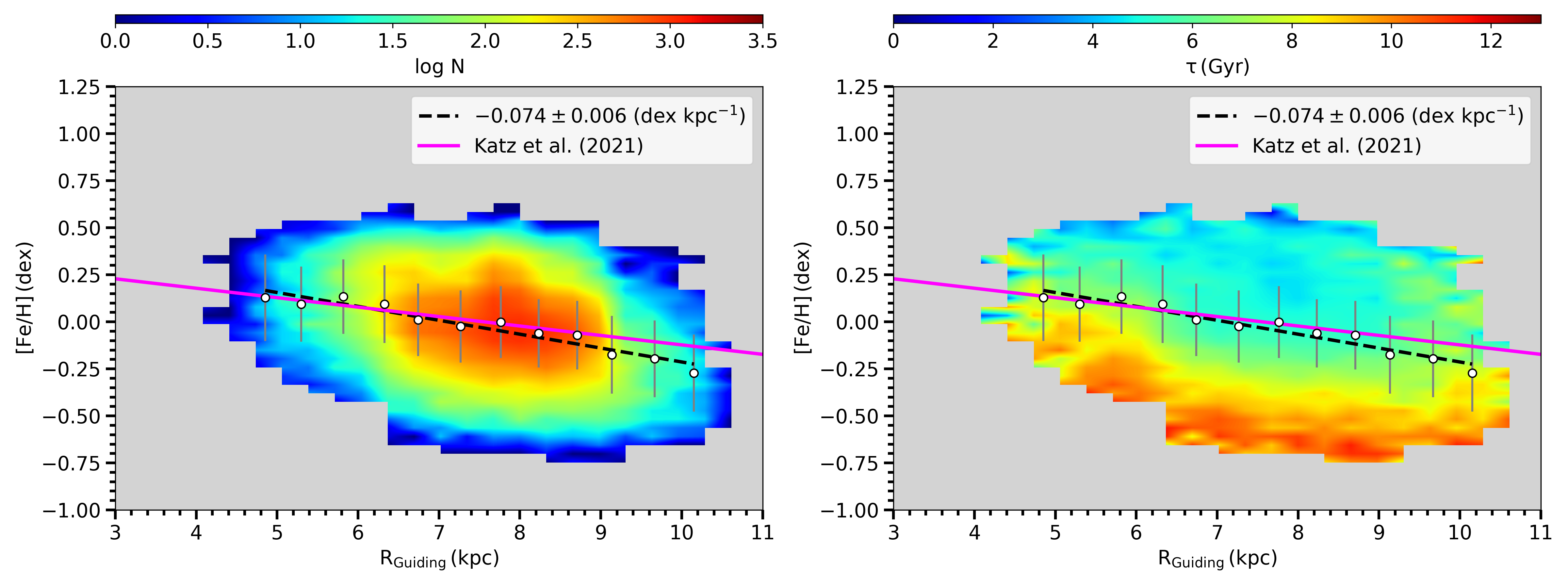}
\includegraphics[width=\textwidth]{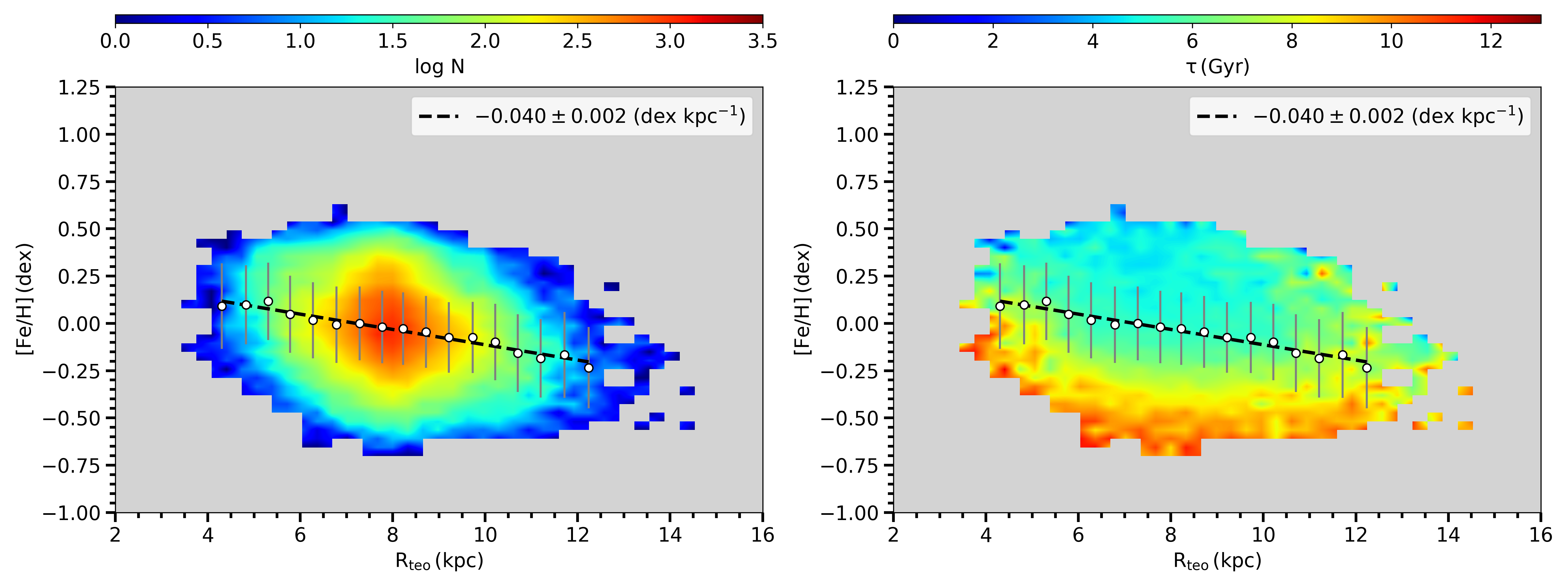}
\caption{Radial metallicity gradients calculated for guiding radii (upper panels) and traceback early orbital radii  (lower panels) of thin disc main-sequence stars. White circles represent the locii of the data and black solid lines represent the linear fit. Colour code for logarithmic number density (left panels) and for stellar age (right panels).} \label{fig:rguifeh}
\end{figure*}

\section{Results and Discussion} \label{sec:discussion}

\subsection{Radial Elemental Gradients}
The radial metallicity gradients are calculated for [Fe/H], [$\alpha$/Fe] and [Mg/H] abundance ratios using the guiding and traceback early orbital radius calculated from the axisymmetrical potential for the Galactic thin disc. The analysis on iron gradients is shown in Figure~\ref{fig:rguifeh} with two colour-coded panels. The left panels of Figure ~\ref{fig:rguifeh} represent logarithmic number density, while the right panels represent stellar age. In this figure, the white points represent the median values of the binned data, with the vertical error bars indicating the standard deviation within each bin. These bins were constructed using distance intervals of 500 pc, ensuring that each bin contains a minimum of 100 data points. The gradient was calculated using the least squares method to ensure the robustness of the statistical analysis. Based on the calculations of the fit function obtained for the loci of the data, the radial metallicity gradient was found to be $d{\rm [Fe/H]}/dR_{\rm Guiding}=-0.074\pm 0.006$ dex kpc$^{-1}$ and $d{\rm [Fe/H]}/dR_{\rm teo}=-0.040\pm 0.002$ dex kpc$^{-1}$, indicating a significant gradient for the thin disc population.

\begin{figure*}
\centering
\includegraphics[width=\textwidth]{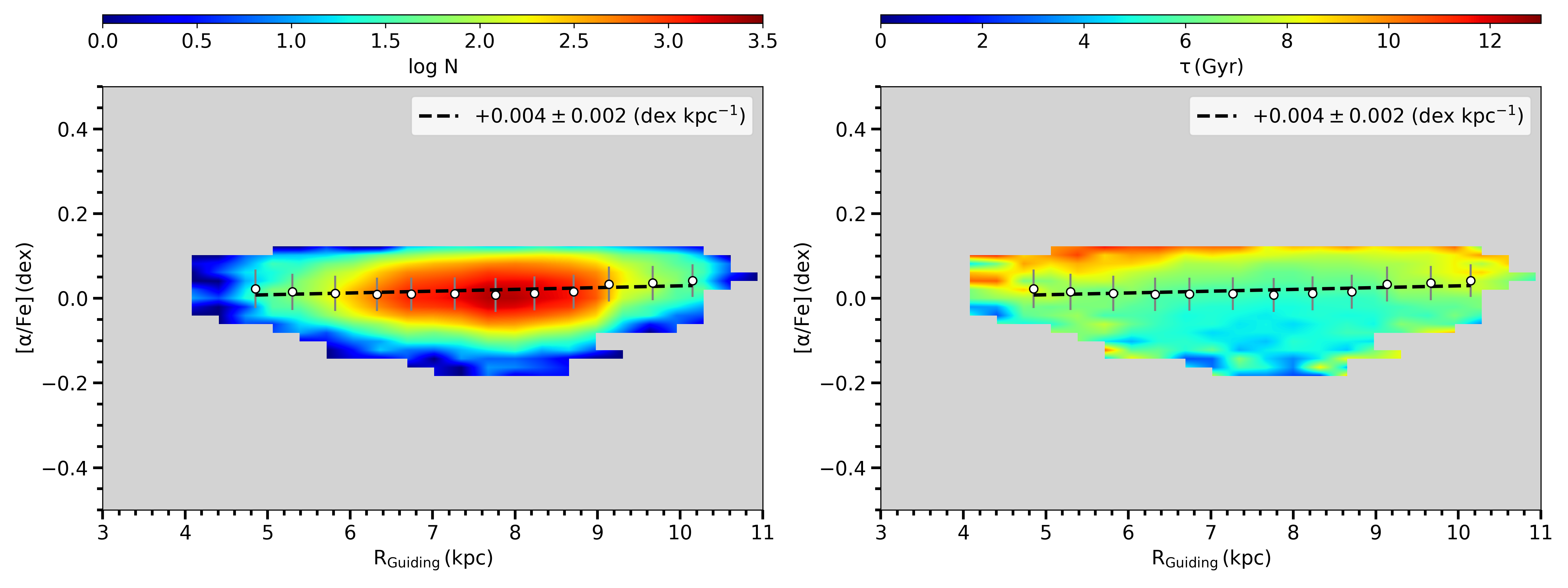}
\includegraphics[width=\textwidth]{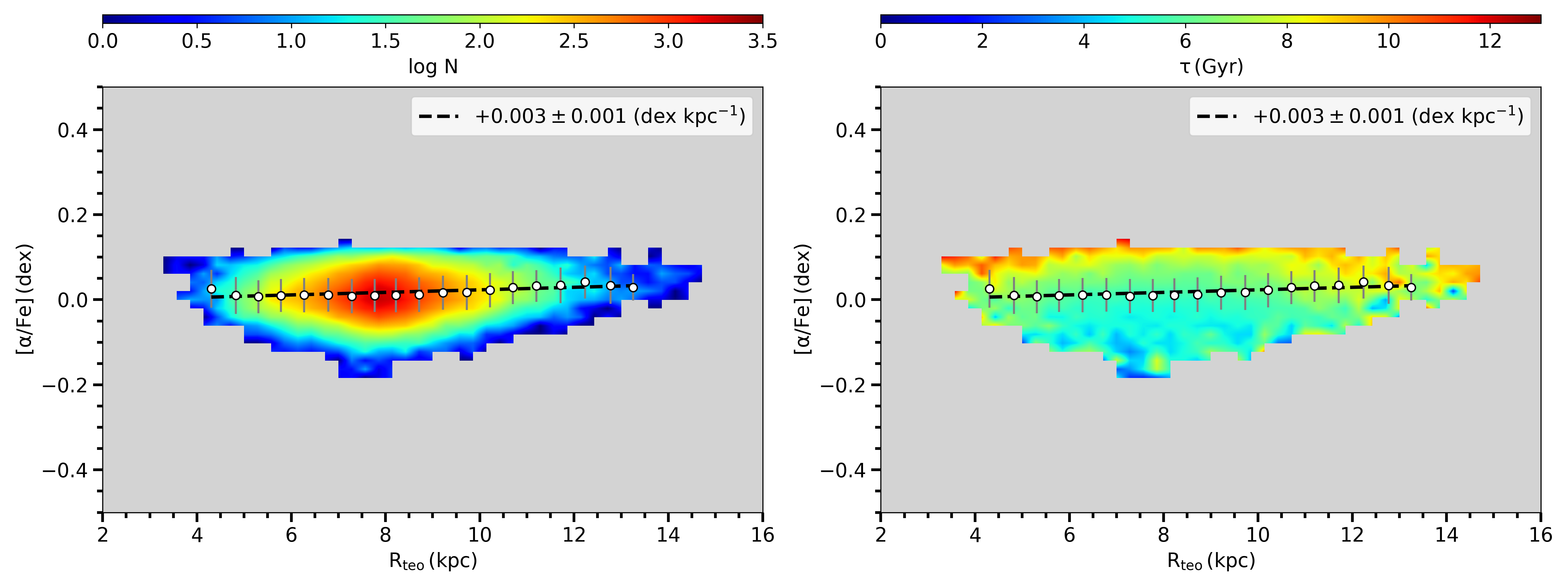}
\caption{Radial ${\alpha}$-elements gradients calculated for guiding radii (upper panels) and traceback early orbital radii (lower panels) of thin disc main-sequence stars. White circles represent the locii of the data and black solid lines represent the linear fit. Colour code for logarithmic number density (a) and for stellar age (b).} \label{fig:rguialpha}
\end{figure*}

\begin{figure*}[ht!]
\centering
\includegraphics[width=\textwidth]{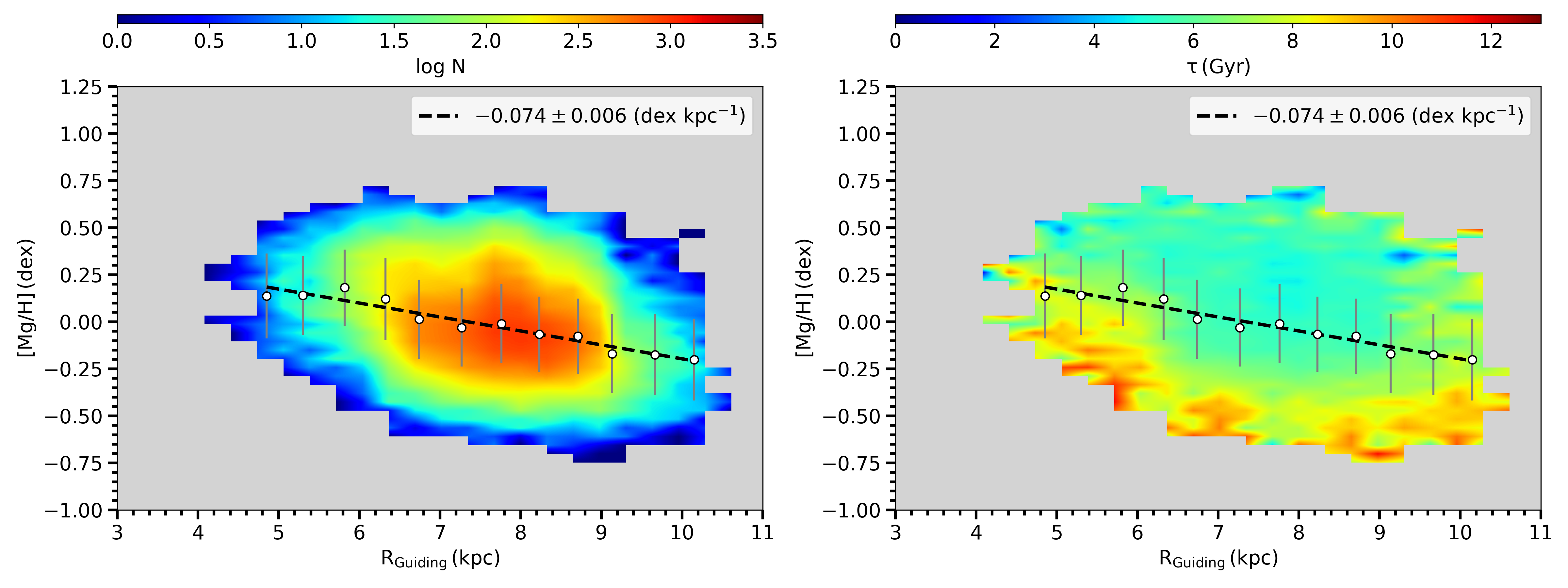}
\includegraphics[width=\textwidth]{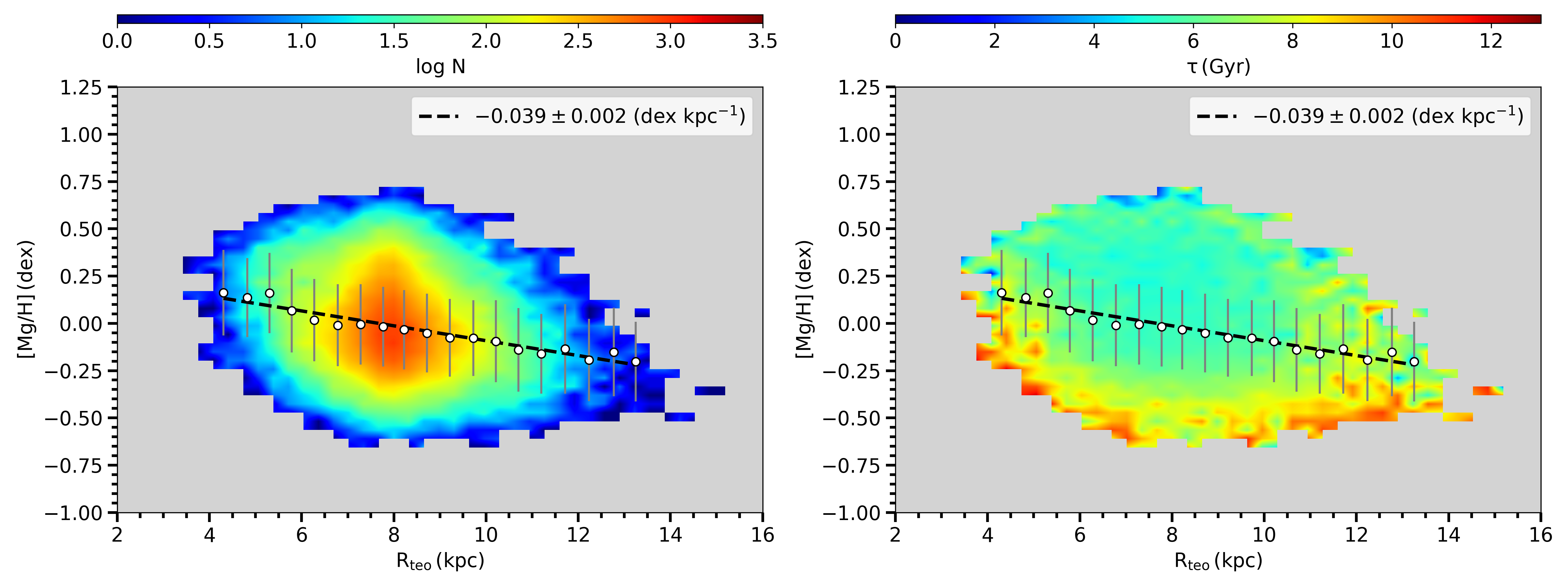}
\caption{Radial magnesium gradients calculated for guiding radii (upper panels) and traceback early orbital radii (lower panels) of thin disc main-sequence stars. White circles represent the locii of the data and black solid lines represent the linear fit. Colour code for logarithmic number density (left panels) and for stellar age (right panels).} \label{fig:rguimgh}
\end{figure*}

Similarly, the analysis of the radial [$\alpha$/Fe] and [Mg/H] abundance gradients are estimated using the guiding radius of the star sample. The resulting gradients for [$\alpha$/Fe] found as $d[\alpha/{\rm Fe]}/dR_{\rm Guiding}=+0.004\pm 0.002$ dex kpc$^{-1}$, and  $d[\alpha/{\rm Fe]}/dR_{\rm teo}=+0.003\pm 0.001$ dex kpc$^{-1}$ and is shown in Figure~\ref{fig:rguialpha} in the same fashion as in Figure~\ref{fig:rguifeh}. This result suggests that there is no prominent $\alpha$-abundance gradient for the thin disc main-sequence stars. Furthermore, the radial gradients for [Mg/H] are found to be $d{\rm [Mg/H]}/dR_{\rm Guiding}=-0.074\pm 0.006$ dex kpc$^{-1}$ for the guiding radius and  $d{\rm [Mg/H]}/dR_{\rm teo}=-0.039\pm 0.002$ dex kpc$^{-1}$ for the traceback early orbital radius.  These gradients are shown in Figure~\ref{fig:rguimgh} with panels representing both the logarithmic number density and stellar age, respectively.

\subsection{Comparing Radial Metallicity Gradients with Literature}
Overall radial metallicity ([Fe/H]) gradient for $R_{\rm Guiding}$ has given similar results to \citet{Plevne2015}. In the study by \citet{Plevne2015} radial metallicity gradients were calculated for the guiding radius of RAVE DR4 main-sequence stars. Their thin disc main-sequence sample was separated into $0<Z_{\rm max} {\rm (kpc)}<0.5$ and $0.5<Z_{\rm max} {\rm (kpc)}<0.8$ intervals and the metallicity gradients were obtained as $d{\rm [Fe/H]}/dR_{\rm Guiding}=-0.083\pm0.030$ dex kpc$^{-1}$ and $d{\rm [Fe/H]}/dR_{\rm Guiding}=-0.065\pm0.039$ dex kpc$^{-1}$, respectively. Similarly, \citet{Boecheetal2013} obtained $d{\rm [Fe/H]}/dR_{\rm Guiding}=-0.065\pm0.039$ dex kpc$^{-1}$ for 19,962 RAVE main-sequence stars within the $0 < Z_{\rm max}$ {\rm (kpc)} $\leq$ 0.4 interval. These findings are in good agreement with the result of $d[{\rm Fe/H}]/dR_{\rm Guiding}=-0.074\pm 0.006$ dex kpc$^{-1}$ from this study. The radial gradients of [$\alpha$/Fe] over $R_{\rm Guiding}$ for the thin-disc stars were found to be +0.004$\pm$0.002 dex kpc$^{-1}$. This gradient agrees with the \citet{Liangetal2019} main-sequence sample compiled from APOGEE and LAMOST sky surveys. Additionally, the radial metallicity gradients obtained in this study are in good agreement with the metal and $\alpha$-element gradients $-0.056\pm0.007$ and $+0.007\pm0.001$ dex kpc$^{-1}$, respectively, given for stars of different types and at different Galactocentric distances spectroscopically analyzed in the {\it Gaia} survey \citep{GaiaDR3-2023b}. It turns out that radial metallicity gradient values under different constraints have produced results in agreement with the literature studies. Using a homogeneous sample of main-sequence stars with precise spectroscopic and astrometric data has provided reliable metallicity gradients. 

\citet{Hou2000} have provided observational values for the magnesium gradient as -0.070$\pm$0.001 dex kpc$^{-1}$, which is in good aggrement with this study. \cite{MarsakovBarkova2004}, which investigate 77 F-G main-seqeunce stars in the Solar neighbourhood, found a radial [Mg/H] gradient as -0.026$\pm$0.007 dex kpc$^{-1}$. \cite{Bergemann2014} have studied FGK spectral type disc stars using Gaia-ESO large stellar survey data. In this study, they found a radial magnesium gradient for stars in $|Z|\leq 0.3$ kpc as 0.021$\pm$0.014 dex kpc$^{-1}$, while for stars in $0.3<|Z|{\rm (kpc)} \leq0.8$ range as -0.045$\pm$0.011 dex kpc$^{-1}$. Also, radial magnesium gradients have been estimated for stellar age. Based on this analysis, they found the magnesium gradient as 0.015$\pm$0.014 dex kpc$^{-1}$ for $\tau\leq$ 7 Gyr and as -0.071$\pm$0.029 dex kpc$^{-1}$ for $\tau>12$ Gyr. As can be seen from the studies in the literature, the results obtained for the magnesium gradient calculated from both the guiding and the birth radii are found to be quite compatible with those found in this study.

Figure \ref{fig:rguifeh} shows the comparison of the radial metallicity gradient with the results of \citet{Katezetal2021}. Our analysis, shown by the black dashed line, reveals a gradient of $-0.074 \pm 0.006$ dex kpc$^{-1}$, which is consistent with \citet{Katezetal2021}'s findings (represented by the pink line) within the boundaries of our data set. \citet{Katezetal2021} used APOGEE data, and their results align well with our findings, confirming the robustness of our gradient calculations.

\subsection{Comparing Radial Metallicity Gradients with Galactic Chemical Evolution Model}

In this study, we utilized the chemical evolution model employed in the \citet{Spitonietal2021} study to assess the observed radial [Fe/H] abundance values obtained. The model is an extension of a previous model \citep[e.g.,][]{MatteucciFrancois1989, Spitonietal2019, Spitonietal2020} designed for the Solar neighbourhood, and it is applied to different Galactocentric regions centered at 4 kpc, 8 kpc, and 12 kpc. The model assumes that the Milky Way disc has been formed by two distinct accretion episodes of gas and that the gas infall rate is a function of the Galactic distance $R$ \citep[Equation 1]{Spitonietal2021}. The gas infall rate is described by a set of equations that take into account the timescales of gas accretion for the formation of the high-$\alpha$ and low-$\alpha$ disc phases. The star formation rate is described by a set of equations that considered the star formation efficiency and gas consumption timescale, which are functions of the Galactocentric distance $R$ \citep{Kennicutt1998}. The authors adopt the nucleosynthesis yields from various sources: \citet{vandenHoekGroenewegen1997} for single low- and intermediate-mass stars, \citet{Iwamoto1999}  for Type Ia supernovae (using the W7 model), and \citet{WoosleyWeaver1995} for massive stars. For further details on these yields and their applications, see \citet{Spitonietal2021} and \citet{Francoisetal2004}. The researchers use Markov Chain Monte Carlo (MCMC) methods to fit the free parameters of the model to the observed abundance ratios of stars in the APOGEE DR16 sample \citep{ApogeeDR16} at different Galactocentric distances. The model can reproduce the observed trends in [Mg/Fe] and [Fe/H] abundance ratios and explores the evolution of the Galactic disc over time. 

We calculated the radial abundance gradient ($d[{\rm Fe/H}]/dR$) suggested by the model and its time-dependent evolution. To compute the suggested radial metallicity gradient of the model, we obtained \citet{Spitonietal2021}'s model predictions for different Galactic radii and different ages through private communication with E. Spitoni. We modelled the [Fe/H] variation in the Galactic disc for three different radial regions (4, 8, 12 kpc) throughout 13.6 Gyr. As a result of these calculations, the radial [Fe/H] gradient suggested by the model for the present day was determined to be $-0.068\pm0.010$ dex kpc$^{-1}$. This value is consistent with the value of $-0.074\pm 0.006$ dex kpc$^{-1}$ obtained in this study.

\begin{figure*}[!ht]
\centering
\includegraphics[width=\textwidth]{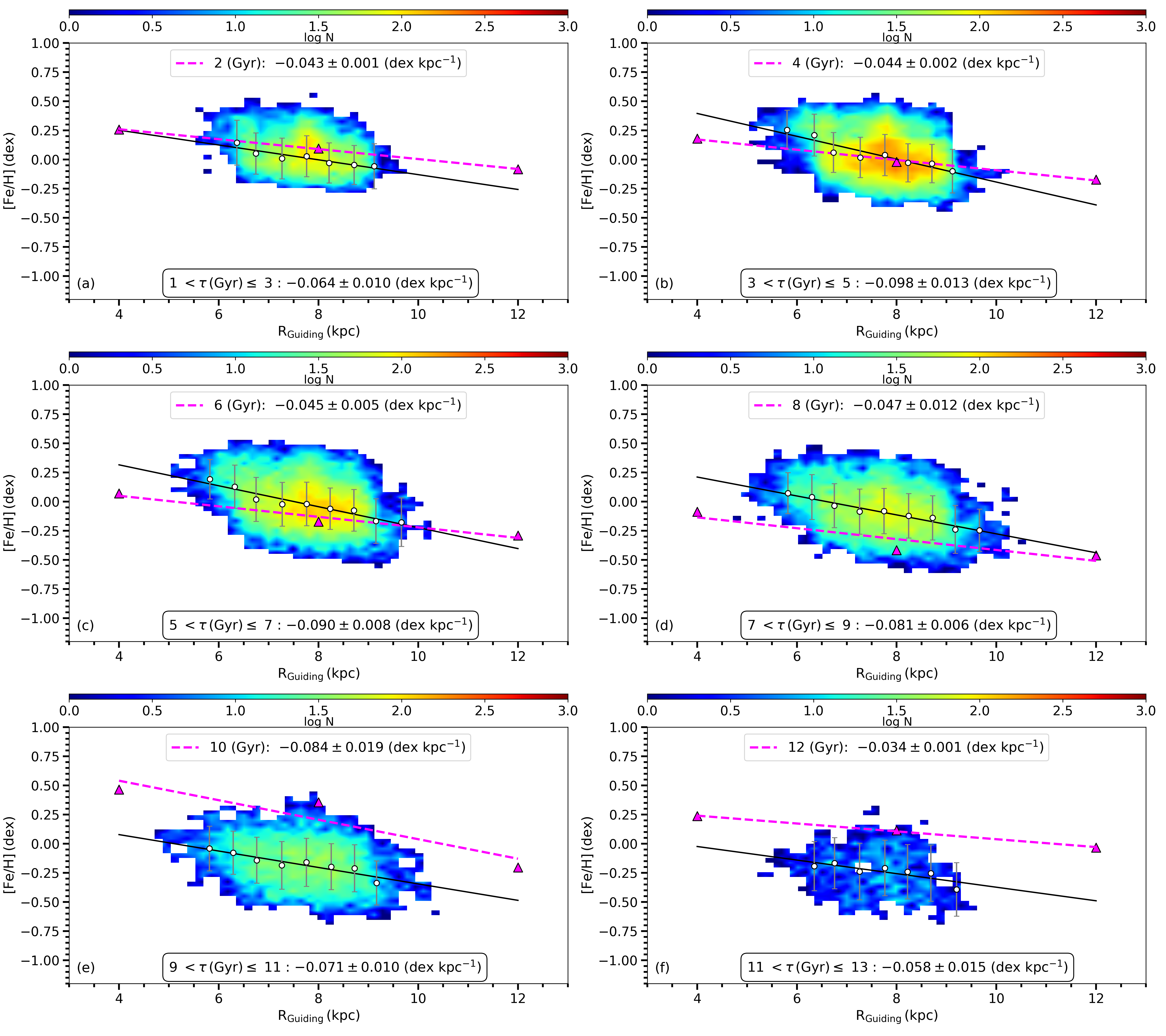}
\caption{Comparison of the radial metallicity gradients for different age intervals in this study with the ones calculated by chemical evolution model of \citet{Spitonietal2021}.}\label{fig:Spitoni}
\end{figure*}

As a result, it has been shown that the radial metallicity gradients calculated from the current and precise data of the main-sequence stars within the Solar neighbourhood are compatible with those produced from the theoretical models. This study demonstrates the necessity of testing the theoretical models produced for regions beyond the Sun by applying them to other star groups with different luminosity classes.

\subsection{Contribution of Results to Models}
Gradient profiles of the measured radial abundance of elements, obtained through observational studies, constitute one of the most critical empirical findings that have contributed to the emergence of Galactic chemical evolution models. Consequently, they are still widely employed for testing their predictions today. While observational studies measure the current gradient value, it is crucial to understand the history of the Galaxy to determine whether this gradient has remained the same from the past or how it has changed. Therefore, in this study, we divided our sample into age groups calculated using BSTEP ages \citep{Sharmaetal2018} and computed the radial iron abundance gradient for each age group. Simultaneously, we calculated the predicted gradients for each age group according to the chemical evolution model \citep{Spitonietal2021}. The calculated observational and model-predicted gradients are depicted in Figure~\ref{fig:Spitoni}. As seen from the figure, while the metallicity gradients found by the model and observational data are consistent with each other for the 2-4-6 Gyr age groups, they diverge in terms of slope for the 8-10-12 Gyr groups, although showing consistent trends in terms of suggested mean abundance. Particularly, the model proposes a significantly richer environment during these periods corresponding to the early stages of the Milky Way, compared to the observational data. One potential reason for this discrepancy could be the model predicting a more enriched environment than the observational data. However, it's important to note that this study relies on observational data of main-sequence stars in the Solar neighbourhood. Figure~\ref{fig:Spitoni} shows that observational data at different ages and model data are in good agreement up to 8 Gyrs. However, in the diagrams plotted for the ages of 10 and 12 Gyr, it was found that the model data were extremely metal-rich compared to the younger ages and did not provide the observational data.

\subsection{Radial Orbital Variation}

Radial orbital variation emerges as an integral component of studies on the Galactic metallicity gradient. Throughout their lifetimes, stars move in their orbits around the center of the Galaxy and are subjected to the influences of various perturbation sources within the Galaxy. Depending on their interactions with these perturbation sources, stars may drift away from their birth radius \citep{Lu2022}. Given that radial orbital variation directly affects the results of gradient studies, it is crucial to quantify this impact. Thus, determining a star’s traceback early orbital radius is of significant importance. In this study, the $R_{\rm teo}$ of stars were calculated using the method described in Section 3.1.1.

\begin{figure}
    \centering
    \includegraphics[width=\columnwidth]{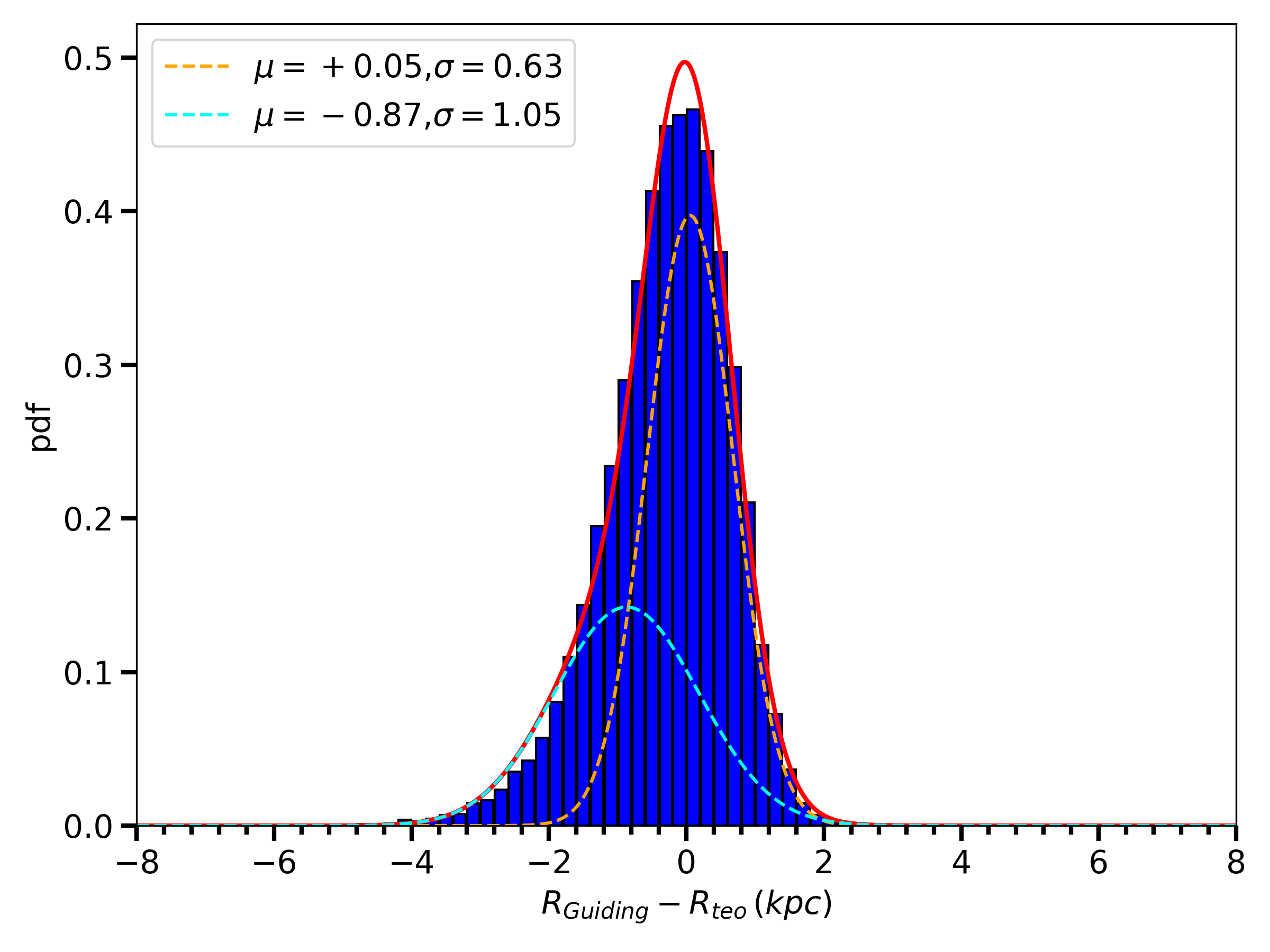}
    \caption{Distribution of the difference between guiding radii and traceback early orbital radii ($R_{\text{Guiding}} - R_{\text{teo}}$). The histogram is modelled with two Gaussian distributions: one with $\mu = +0.05$, $\sigma = 0.63$ (orange dashed line) and the other with $\mu = -0.87$, $\sigma = 1.05$ (blue dashed line).}
    \label{fig:migration_gaussian_fit}
\end{figure}

To understand the extent of radial orbital variation in the sample, we analyzed the difference between the $R_{\text{Guiding}}$ and $R_{\text{teo}}$ values of the stars. The distribution of this difference is shown in Figure~\ref{fig:migration_gaussian_fit}. Upon traceback early orbital radius analysis, the skewness of the distribution was calculated to be -1.60, indicating that it could not be represented by a single symmetric distribution. Consequently, two separate normal distributions were fitted to represent stars influenced by radial orbital variation and those not influenced by it. These distributions were found to be $\mathcal{N}(-0.87, 1.05^2)$ and $\mathcal{N}(0.05, 0.63^2)$, respectively.

\citet{Kubryketal2015a} defined the effect of radial migration as a star moving at least 2 kpc away from its birth radius. Considering this definition, the distribution representing stars not influenced by radial orbital variation, shown as the orange dashed line in Figure \ref{fig:migration_gaussian_fit}, centers between 0.8 and 1.2 kpc. This suggests that the distributions accurately represent their respective populations. Stars affected by radial orbital variation are represented by a much broader distribution. The total effect of radial orbital variation in the sample was determined by integrating the difference between the areas covered by these two distributions relative to the total sample area, resulting in a maximum of 6\%. This indicates that up to 6\% of the thin disc stars in the sample might be affected by radial orbital variation.

\subsubsection{Radial Orbital Variation Stellar Age Relation}

\begin{figure*}
    \centering
    \includegraphics[width=0.96\textwidth]{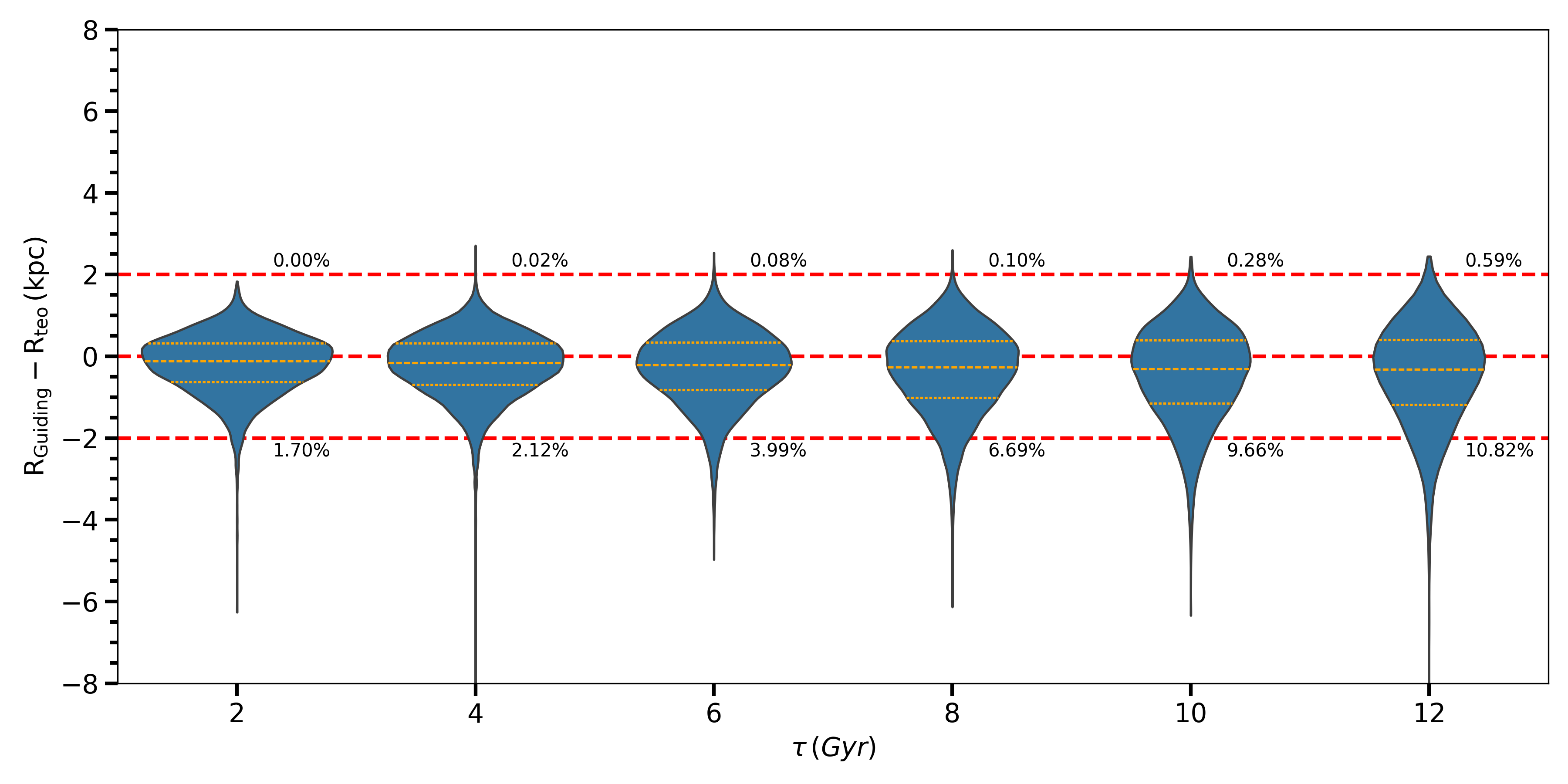}
    \caption{Violin plot showing the distribution of the difference between guiding radii and traceback early orbital radii ($R_{\text{Guiding}} - R_{\text{teo}}$) versus stellar ages ($\tau$). Red dashed lines indicate orbital variation contamination thresholds, and percentages denote the proportion of stars beyond these limits. The orange dashed and dotted lines in each age interval show the median and standard deviations in the subsample, respectively.}
    \label{fig:R_guiding_birth_diff_age_violin}
\end{figure*}

\begin{figure*}
    \centering
    \includegraphics[width=0.96\textwidth]{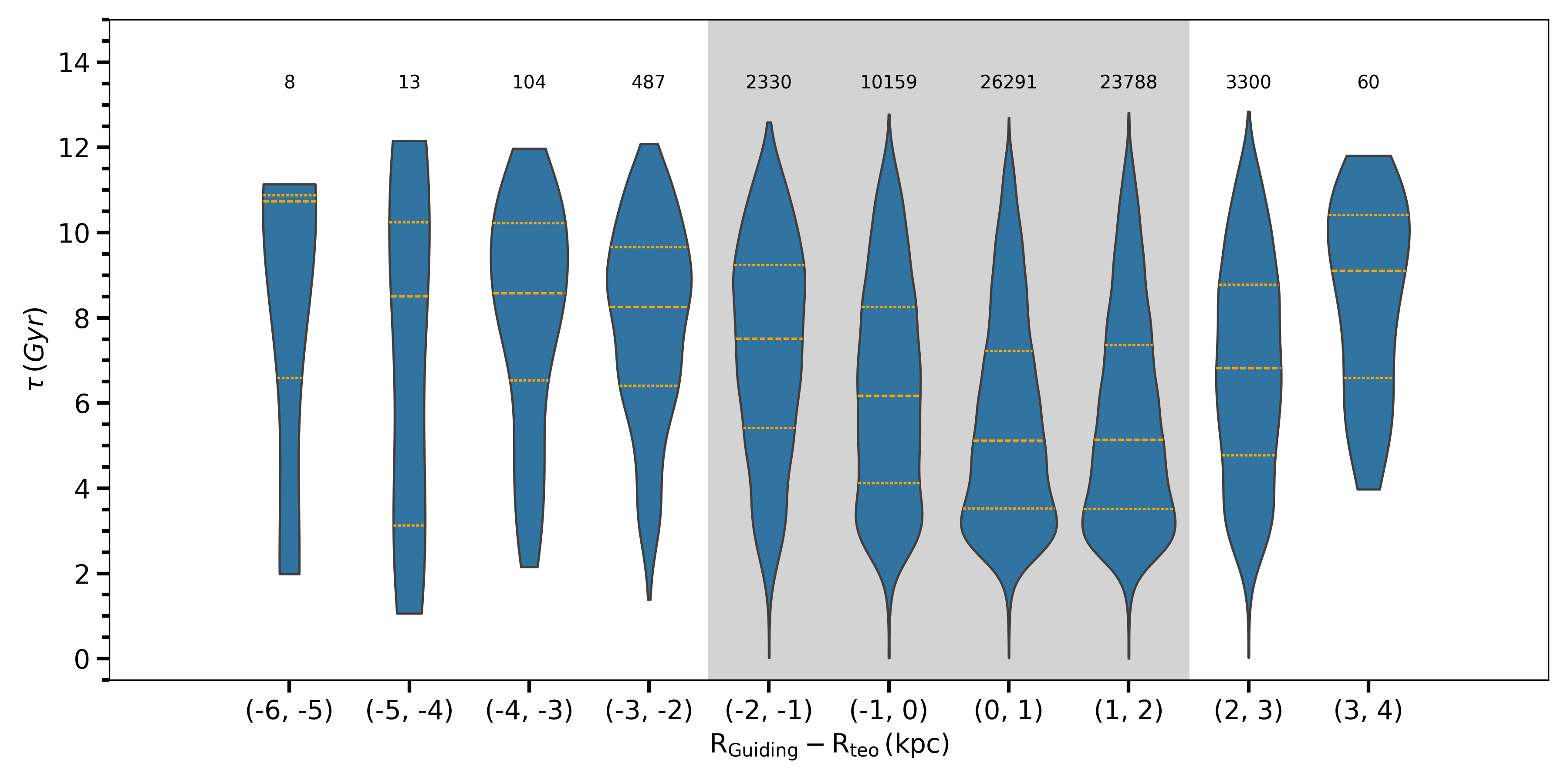}
    \caption{Violin plot showing the distribution of stellar ages ($\tau$) versus the difference between guiding radii and traceback early orbital radii ($R_{\text{Guiding}} - R_{\text{teo}}$). The shaded region represents the range where the stars' guiding radii are close to their traceback early orbital radii. The dashed and dotted lines in each panel show the median and standard deviations in the subsample, respectively.}
    \label{fig:R_guiding_birth_diff_age_violin_step}
\end{figure*}

After identifying that the sample contains stars affected by radial orbital variation, we investigated the age dependency of this radial orbital variation effect. Figure~\ref{fig:R_guiding_birth_diff_age_violin} shows the violin plot which represents the distribution of $R_{\rm Guiding}-R_{\rm teo}$ versus $\tau$. The red dashed line in Figure~\ref{fig:R_guiding_birth_diff_age_violin} represents the $\pm$2 kpc radial orbital variation threshold accepted in the study. The orange dashed lines within the distributions indicate the 16\%, 50\%, and 84\%. Figure~\ref{fig:R_guiding_birth_diff_age_violin} inspection reveals that the proportion of stars experiencing radial orbital variation increases with mean age. Additionally, it is evident that the majority of the sample originates from a radius larger than the Solar neighbourhood, indicating that the radial orbital variation mechanism predominantly occurs from the outer regions inward in this sample. The proportional increase in radial orbital variation with age can be attributed to the long-term effect on older stars from perturbation sources. Furthermore, Figure~\ref{fig:R_guiding_birth_diff_age_violin} shows that the percentiles of the age-dependent distribution remain relatively unchanged. While the tails of the distributions extend, the central values remain almost constant, suggesting that the characteristic distributions of the sub-samples are not significantly affected by radial orbital variation.

Figure~\ref{fig:R_guiding_birth_diff_age_violin_step} represents a violin plot among $\tau$ versus $R_{\rm {Guiding}} - R_{\rm {teo}}$. The gray region on the diagram shows where the radial orbital variation has no efficiency ($|R_{\rm {Guiding}} - R_{\rm {teo}}|\leq 2$ kpc). Upon inspection, it can be seen that the subgroups not affected by radial orbital variation have similar median values and distributions for their mean ages. In contrast, those outside the gray region are relatively older and have a more varied distribution. This observation further indicates that the effect of radial orbital variation becomes more pronounced as the age of the sample increases.

\subsection{The Impact of Radial Orbital Variation on Metallicity Gradients}

We analyze the impact of radial orbital variation on metallicity gradients in the Galactic disc, focusing on how these gradients vary across different age ranges. Table~\ref{tab:metallicity_gradients} presents the radial metallicity gradients ($d{\rm [Fe/H]}/dR_{\text{Guiding}}$) for various age intervals, comparing uncorrected and corrected values. 

Taking into account all main-sequence stars, the calculated radial metal metallicity gradient is -0.074$\pm$0.005 dex kpc$^{-1}$, whereas the radial metallicity gradient calculated by excluding radial migrated stars ($\sim$ 6\%) is -0.068$\pm$0.006 dex kpc$^{-1}$. The results indicate changes due to radial orbital variation, particularly in older stars. For younger stars (1-3 Gyr), the uncorrected metallicity gradient is -0.064$\pm$0.010 dex kpc$^{-1}$, and the corrected gradient is -0.057$\pm$0.010 dex kpc$^{-1}$, with a radial orbital variation contamination of 1.70\%. This low contamination suggests minimal radial orbital variation, preserving original chemical signatures. In older stars (11-13 Gyr), the uncorrected gradient is -0.058$\pm$0.015 dex kpc$^{-1}$, while the corrected gradient is -0.048$\pm$0.012 dex kpc$^{-1}$, showing a radial orbital variation contamination of 11.41\%. This indicates that while there is some effect of radial orbital variation, the impact is not overwhelmingly significant due to the relatively small number of older stars in the sample. This demographic distribution allows for a clearer understanding of the younger stars’ original metallicity gradients without significant interference from radial orbital variation.

In contrast, older stars exhibit a broader distribution of $R_{\text{Guiding}} - R_{\text{teo}}$ , indicating substantial radial orbital variation over time. This broader distribution results in more complex interpretations of metallicity gradients, as the chemical signatures of these stars have been altered by their radial orbital variation from their traceback early orbital radii. It is important to note that the minimal impact of radial orbital variation on the results aligns with some studies in the literature. For instance, \citet{Grand2015} has found similar results, indicating that when the effect of radial orbital variation is minimal, the metallicity gradients are more accurate and reliable. This study also shows that the impact of radial orbital variation on the results is minimal, allowing for a clearer and more accurate interpretation of the observed metallicity gradients. 


\section{Summary and Conclusion}

\begin{table*}[]
\caption{Radial metallicity gradients calculated according to the presence or absence of radial orbital variation according to the age groups of the stars in the sample.} \label{tab:metallicity_gradients}
\resizebox{\textwidth}{!}{%
\begin{tabular}{c|rc|rc|c|}
\cline{2-6}
                                           & \multicolumn{2}{c|}{Radial Orbital Variation Uncorrected}       & \multicolumn{2}{c|}{Radial Orbital Variation Corrected}         & \multirow{2}{*}{\begin{tabular}[c]{@{}c@{}}Radial Orbital \\ Variation Contamination\end{tabular}} \\ \cline{1-5}
\multicolumn{1}{|c|}{$\tau$}               & \multicolumn{1}{c|}{$N$}     & $d{\rm [Fe/H]}/dR_{\rm Guiding}$ & \multicolumn{1}{c|}{$N$}     & $d{\rm [Fe/H]}/dR_{\rm Guiding}$ &                                                                                                    \\ \hline
\multicolumn{1}{|c|}{Gyr}                  & \multicolumn{1}{c|}{}        & ${\rm(dex}$ ${\rm kpc^{-1})}$    & \multicolumn{1}{c|}{}        & ${\rm(dex}$ ${\rm kpc^{-1})}$    & \%                                                                                                 \\ \hline
\multicolumn{1}{|c|}{1 $ < \tau \leq$ 3}   & \multicolumn{1}{r|}{8\,749}  & -0.064 $\pm$ 0.010               & \multicolumn{1}{r|}{8\,600}  & -0.057 $\pm$ 0.010               & 1.70                                                                                               \\ \hline
\multicolumn{1}{|c|}{3 $ < \tau \leq$ 5}   & \multicolumn{1}{r|}{20\,228} & -0.098 $\pm$ 0.013               & \multicolumn{1}{r|}{19\,796} & -0.099 $\pm$ 0.014               & 2.14                                                                                               \\ \hline
\multicolumn{1}{|c|}{5 $ < \tau \leq$ 7}   & \multicolumn{1}{r|}{16\,134} & -0.090 $\pm$ 0.008               & \multicolumn{1}{r|}{15\,477} & -0.097 $\pm$ 0.010               & 4.07                                                                                               \\ \hline
\multicolumn{1}{|c|}{7 $ < \tau \leq$ 9}   & \multicolumn{1}{r|}{11\,999} & -0.081 $\pm$ 0.006               & \multicolumn{1}{r|}{11\,184} & -0.084 $\pm$ 0.007               & 6.79                                                                                               \\ \hline
\multicolumn{1}{|c|}{9 $ < \tau \leq$ 11}  & \multicolumn{1}{r|}{7\,617}  & -0.071 $\pm$ 0.010               & \multicolumn{1}{r|}{6\,860}  & -0.075 $\pm$ 0.010               & 9.94                                                                                               \\ \hline
\multicolumn{1}{|c|}{11 $ < \tau \leq$ 13} & \multicolumn{1}{r|}{1\,701}  & -0.058 $\pm$ 0.015               & \multicolumn{1}{r|}{1\,507}  & -0.048 $\pm$ 0.012               & 11.41                                                                                              \\ \hline
\multicolumn{1}{|c|}{$0<\tau \leq$ 13.8}   & \multicolumn{1}{r|}{66\,545} & -0.074 $\pm$ 0.005               & \multicolumn{1}{r|}{62\,568} & -0.068 $\pm$ 0.006               & 5.97                                                                                               \\ \hline
\end{tabular}%
}
\end{table*}

We have investigated the radial metallicity gradients for the chemically determined thin-disc population using the dynamic orbital parameters of main-sequence stars selected from the GALAH DR3 catalogue. Astrometric data is provided within the value-added catalogue of GALAH DR3 for the sample stars along with ages calculated with the BSTEP algorithm. Galactic populations determined on the $[\alpha/{\rm Fe}]\times$[Fe/H] plane using GMM. Stellar orbital parameters are obtained for axisymmetric Galactic potential, {\sc MWPotential2014} model, using {\sc Galpy} \citep{galpy}. The radial metallicity gradients for [Fe/H], [$\alpha$/Fe] and [Mg/H] abundance ratios are calculated for the guiding and traceback early orbital radius of thin disc main-sequence stars as a function of both number density and stellar age. Radial abundance gradients are determined via linear fits applied on the locii points on each plane.

When the time-dependent variation of the radial metallicity gradient is examined, it is observed that observational findings indicate the presence of a gradient along the disc throughout time, maintaining negative values from the early stages of the Galaxy to the present day. Similarly, when the model is investigated, a similar prediction is observed. The classical two-infall approach explains this negative abundance gradient observational finding by assuming that the disc has collapsed from inside to outside. However, the delayed collapse time approach of the model we compare in the study does not rely on this assumption. Nevertheless, it still predicts a negative gradient consistently with observational findings. Therefore, the model demonstrates that without a prior assumption, it performs better in estimating the observational findings than its classical version. The fact that it predicts this without assuming a time-dependent change in gradient and an inside-out formation assumption demonstrates that inside-out collapse is not an indispensable assumption for the gradient.

The study demonstrates that radial orbital variation becomes more pronounced with the age of the stellar population, as clearly indicated in Table~\ref{tab:metallicity_gradients}. This trend can be attributed to the prolonged exposure of older stars to the physical processes within the Galaxy that drive radial orbital variation. Considering the age-dependent metallicity gradient changes in Table~\ref{tab:metallicity_gradients}, when excluding the 1-3 Gyr interval, we observe a clear trend where the metallicity gradient becomes increasingly negative from the older stars (11-13 Gyr) to the younger stars. Although the 1-3 Gyr data point does not follow this trend precisely, the remaining younger age groups still exhibit a more negative gradient compared to the oldest sample. This trend reflects how the chemical evolution of the Milky Way has resulted in steeper gradients for more recently formed stars. The deviation of the 1-3 Gyr data point from the overall trend may be due to its limited representation of the gradient, as seen in Figure~\ref{fig:Spitoni}(a). The data for this interval may not cover a sufficiently wide range to accurately represent the true gradient. However, given the uncertainties involved, this variation in the gradient should not be considered statistically significant.

The impact of radial orbital variation on the radial metallicity gradient and its temporal evolution is further analyzed in Table~\ref{tab:metallicity_gradients}. Although the effect of radial orbital variation increases with time, its influence on the radial metallicity gradient remains statistically insignificant. This indicates that the selection criteria used for the sample effectively minimized the impact of radial orbital variation on the calculated gradient, thereby ensuring the robustness of the results. Given in Table~\ref{tab:metallicity_gradients}, it is crucial to consider age uncertainties to meaningfully interpret the results. To investigate this, we repeated the analysis using the same steps but limited the sample to those with age uncertainties less than 2 Gyr within each age interval. The results show that the radial metallicity gradient outcomes are not affected by this restriction.

In this study, the star sample has been confined to a limited area within the thin disc due to the utilization of main-sequence stars. Consequently, the results may be limited in terms of generalisability. However, due to the sample being confined to a narrow spatial region, the chemical evolution model could not be applied at different radii; hence, it was compared with the results of other studies. In our forthcoming research, we plan to reevaluate these findings by applying the model to the sample with a larger spatial volume with more precise ages. Through this planned study, we aim to investigate whether the negative element abundance gradient is a result of Galactic chemical evolution or a reflection of a pre-existing phenomenon from the beginning to the present day.

\section*{Acknowledgments}
We thank the anonymous referee and editorial board for their insightful and constructive suggestions, which significantly improved the paper. This study is a part of MSc thesis of Furkan Akbaba. We would also like to thank Dr Spitoni for sharing the outputs of his chemical evolution model for use in our study. The GALAH survey is based on observations made at the Australian Astronomical Observatory, under programmes A/2013B/13, A/2014A/25, A/2015A/19, A/2017A/18. We acknowledge the traditional owners of the land on which the AAT stands, the Gamilaraay people, and pay our respects to elders past and present. This work has made use of data from the European Space Agency (ESA) mission {\it Gaia} (\url{https://www.cosmos.esa.int/gaia}), processed by the {\it Gaia}Data Processing and Analysis Consortium (DPAC, \url{https://www.cosmos.esa.int/web/gaia/dpac/consortium}). Funding for the DPAC has been provided by national institutions, in particular, the institutions participating in the {\it Gaia} Multilateral Agreement. This research has made use of NASA’s Astrophysics Data System Bibliographic Services and VizieR catalogue access tool, CDS, Strasbourg, France. 

The following software and programming languages made this research possible: Python \citet{Python}; Astropy \citet{Astropy2013, Astropy2018}; Pandas \citet{Pandas}; NumPy \citet{Numpy}; Matplotlib \citet{Matplotlib}.

\subsection*{Author contributions}

\textbf{Conception/Design of study}: Furkan Akbaba, Tansel Ak, Sel\c cuk Bilir, Olcay Plevne;\\ 
\textbf{Data Acquisition}: Furkan Akbaba, Olcay Plevne; \\
\textbf{Data Analysis/Interpretation}: Furkan Akbaba, Olcay Plevne; \\
\textbf{Drafting Manuscript}: Furkan Akbaba, Tansel Ak, Sel\c cuk Bilir, Olcay Plevne, \"{O}zgecan \"{O}nal Ta\c s, George Seabroke; \\
\textbf{Critical Revision of Manuscript}: Furkan Akbaba, Tansel Ak, Sel\c cuk Bilir, Olcay Plevne, \"{O}zgecan \"{O}nal Ta\c s, George Seabroke; \\
\textbf{Final Approval and Accountability}: Furkan Akbaba, Tansel Ak, Sel\c cuk Bilir, Olcay Plevne, \"{O}zgecan \"{O}nal Ta\c s, George Seabroke.

\subsection*{Financial disclosure}

None reported.

\subsection*{Conflict of interest}

The authors declare no potential conflict of interests.

\bibliography{references}


\end{document}